\documentclass{article}

\usepackage{arxiv}

\usepackage[utf8]{inputenc} 
\usepackage[T1]{fontenc}    
\usepackage[hidelinks]{hyperref}       
\usepackage{url}            
\usepackage{booktabs}       
\usepackage{amsfonts}       
\usepackage{amsmath}
\usepackage{nicefrac}       
\usepackage{microtype}      
\usepackage{cleveref}       
\usepackage{graphicx}
\usepackage[square,sort,comma,numbers]{natbib}
\usepackage{doi}
\usepackage{bm}
\usepackage{caption}
\usepackage{subcaption}
\usepackage{multirow}
\usepackage{makecell}
\usepackage{appendix}
\usepackage[section]{placeins}

\usepackage{algorithm}
\usepackage{algpseudocode}

\title{Blind Source Separation of Single-Channel Mixtures via Multi-Encoder Autoencoders}

\date{September 1, 2023}

\author{Matthew B. ~Webster\\
	Mellowing Factory Co. Ltd.\\
	708-7 Banpo-dong, Seocho-gu\\
	Seoul, South Korea, 06535 \\
	\texttt{webster@mellowingfactory.com} \\
	\And
	Joonnyong ~Lee\thanks{Corresponding author.}\\
	Mellowing Factory Co. Ltd.\\
	708-7 Banpo-dong, Seocho-gu\\
	Seoul, South Korea, 06535 \\
	\texttt{joon@mellowingfactory.com} \\
}

\renewcommand{\shorttitle}{Blind Source Separation of Single-Channel Mixtures via Multi-Encoder Autoencoders}

\hypersetup{
pdftitle={Blind Source Separation of Single-Channel Mixtures via Multi-Encoder Autoencoders},
pdfsubject={eess.SP},
pdfauthor={Matthew B. ~Webster, Joonnyong ~Lee},
pdfkeywords={Artificial neural networks, Self-supervised learning, Blind source separation, Multi-encoder autoencoders},
}

\begin{document}
\maketitle

\begin{abstract}
The task of blind source separation (BSS) involves separating sources from a mixture without prior knowledge of the sources or the mixing system. Single-channel mixtures and non-linear mixtures are a particularly challenging problem in BSS. In this paper, we propose a novel method for addressing BSS with single-channel non-linear mixtures by leveraging the natural feature subspace specialization ability of multi-encoder autoencoders. During the training phase, our method unmixes the input into the separate encoding spaces of the multi-encoder network and then remixes these representations within the decoder for a reconstruction of the input. Then to perform source inference, we introduce a \emph{novel} encoding masking technique whereby masking out all but one of the encodings enables the decoder to estimate a source signal. To this end, we also introduce a \emph{sparse mixing loss} that encourages sparse remixing of source encodings throughout the decoder and a so-called \emph{zero reconstruction loss} on the decoder for coherent source estimations. To analyze and evaluate our method, we conduct experiments on a toy dataset, designed to demonstrate this property of feature subspace specialization, and with real-world biosignal recordings from a polysomnography sleep study for extracting respiration from electrocardiogram and photoplethysmography signals.
\end{abstract}

\section{Introduction}
Source separation is the process of separating a mixture of sources into their unmixed source components. Blind source separation (BSS) aims to perform source separation without any prior knowledge of the sources or the mixing process and is widely applied in various fields such as biomedical engineering, audio processing, image processing, and telecommunications. Classical approaches to blind source separation such as independent component analysis (ICA) \citep{761722, HYVARINEN2000411, hyvarinen2013independent, BARROS1998173} or non-negative matrix factorization (NMF) \citep{Lee1999, 1661352, 10.1007/11679363_32} are effective at separating sources under the right conditions \citep{kofidis2016blind}. For example, these methods assume linear mixing of sources in their most basic form.

Blind source separation has a long-standing shared connection with deep learning as they both originally take inspiration from biological neural networks \citep{Lee1999, hyvarinen2013independent, bell1995}. Early shallow artificial neural network (ANN) approaches showed promise on the BSS task \citep{articlewoowai2002, articlesolazzimirko, Tan2001NonlinearBS, articlealmeidaluis2003}. With the explosive rise of deep learning, recent approaches to source separation often incorporate deep neural architectures employing supervised learning, semi-supervised learning, or transfer learning \citep{10.1162/neco_a_01217, stoller2018adversarial, pmlr-v119-jayaram20a}. Further, in deep neural network architectures such as transformers with multi-headed attention \citep{vaswani2017attention}, or even the pivotal AlexNet model \citep{krizhevsky2012imagenet}, originally trained with two sparsely connected convolutional encoders, the tendency for separated encoding structures to specialize in feature subspaces has been observed \citep{voita-etal-2019-analyzing, voita-etal-2018-context, tang-etal-2018-self, everett2023protocaps}.

Modern artificial neural networks are proven to be a successful approach for many source separation and BSS tasks in the literature \citep{e24010055, 9826377, DBLP:conf/ismir/StollerED18, Chen2016DeepAN, 9413901, inproceedingschen2020, simon2022, DBLP:journals/corr/abs-2302-11824, liu2020separate, Gandelsman_2019_CVPR, Zou_2020_CVPR, 10.1016/j.neucom.2023.126895}. In this paper, we utilize the natural feature subspace specialization of neural networks with multiple encoder-like structures for the task of BSS. Our proposed method employs a fully self-supervised training process such that the network is given no examples of the sources, and therefore, is fully compatible with the problem statement of BSS in which we do not have any prior knowledge of the sources or the mixing process and only have access to the mixtures for separation. The underdetermined BSS (UBSS) scenario, whereby there are more sources than observed mixtures, is considered to be the most challenging type in the literature \citep{HYVARINEN1999429}. A single-channel mixture would be the minimum number of possible observations and is underdetermined by definition if the number of sources is greater than one. Our proposed method utilizes a convolutional multi-encoder-single-decoder autoencoder that learns to separate sources within the multiple encoding spaces of the network. The network decoder learns to remix the concatenated outputs of the multiple encoders to predict the original input mixture. This is achieved with a simple reconstruction loss between the network’s input and output and two regularization methods for enabling blind source estimation: the \emph{sparse mixing loss} and the \emph{zero reconstruction loss}. The sparse mixing loss aims to keep the weight connections shared by source encodings and subsequent mappings sparse throughout the decoder. After training, source estimations are produced using an encoding masking technique where all but a single encoding are masked with zero arrays and then the resulting output of the decoder is the source estimation associated with the active encoding which is not masked. Further regularization is applied using the zero reconstruction loss on the decoder during the training phase.
\begin{figure}
\centering
    \begin{subfigure}[]{1.0\linewidth}
        \centering
        \includegraphics[width=0.45\textwidth]{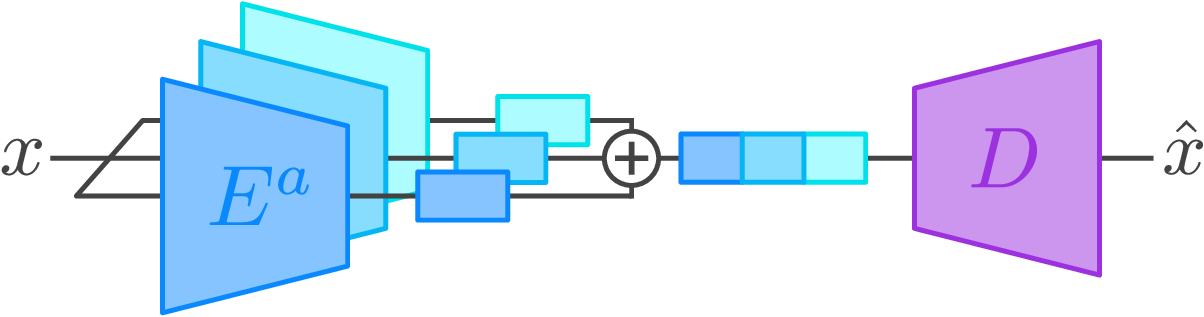}
        \caption{Blind source separation training procedure.}
        \label{fig: Source Separation Training}
    \end{subfigure}

    \begin{subfigure}[]{1.0\linewidth}
        \centering
        \includegraphics[width=0.45\textwidth]{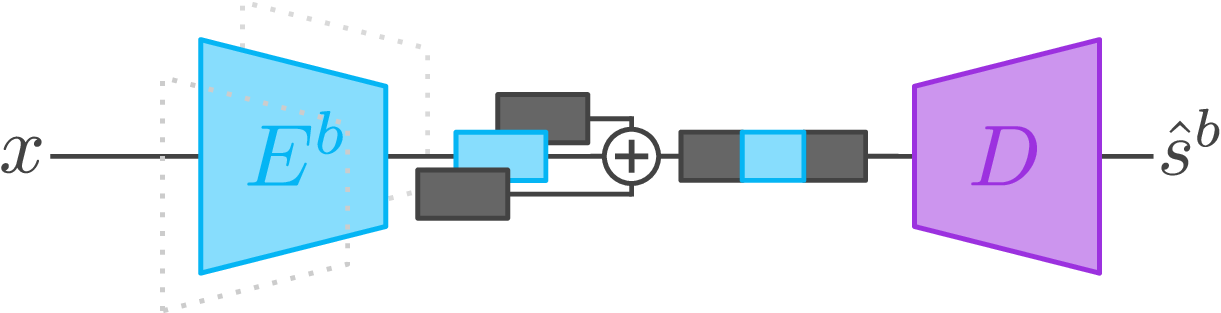}
        \caption{Blind source separation inference procedure.}
        \label{fig: Source Separation Inference}
    \end{subfigure}
    \caption{Illustration of the general training and inference procedure for the proposed method.} 
    \label{fig: Source Separation Procedure}
\end{figure}

To the best of our knowledge, our proposed method is the \emph{first} in the literature to employ fully self-supervised multi-encoder autoencoders for performing BSS. The effectiveness of the proposed method is evaluated on a toy dataset of single-channel post non-linear (PNL) mixtures, and on single-channel real-world electrocardiogram (ECG) and photoplethysmogram (PPG) signals for extracting respiratory signals. The results demonstrate that (a) the encoders of the multi-encoder autoencoder trained with our method can effectively discover and separate the sources of the proposed toy dataset, consisting of post non-linearly mixed triangle and circle shapes, and (b) our method can be successfully applied to estimate a respiratory signal from unprocessed ECG and PPG signals in a fully self-supervised manner with results that outperform existing heuristic and supervised methods. While the proposed method may be applicable to many domains (audio, vision, etc.) and situations for various BSS tasks, this paper focuses on challenging scenarios where learning separation via the distribution of a dataset is a more effective strategy.

We summarize our contributions in this paper as follows:
\begin{itemize}
\item We propose a novel method for BSS by employing fully self-supervised multi-encoder autoencoders for the first time in the literature.
\item We propose a novel encoding masking technique for source separation inference, and two novel model regularization techniques, the \emph{sparse mixing loss} and the \emph{zero reconstruction loss} for implicit source separation learning and source estimation.
\item Our method is the first end-to-end approach in the literature to extract meaningful respiratory signals from unprocessed (no additional filtering or elaborate cleaning) ECG and PPG recordings without supervision from reference respiratory signals or the use of heuristic filtering and algorithms.
\end{itemize}
\section{Related Work}
In the vision domain, deep learning approaches to separating superimposed images, shadow and shading separation, or reflection separation are areas of active research within the domain of computer vision. One work \citep{8682375} approaches the task of separating superimposed handwritten digits from the MNIST dataset and superimposed images of handbags and shoes by learning a single mask to separate two sources. For the first source, the mixed signal is multiplied by the mask, and for the second source, the mixed signal is multiplied by one minus the mask, which inverts the values of the mask. In another work \citep{9616154}, a self-supervised BSS approach based on variational autoencoders makes the assumption that the summation of the extracted sources is equal to the mixed signal i.e. a linear mixture. The authors demonstrated the effectiveness of their approach for separating linearly mixed handwritten digits and for separating mixed audio spectrograms where two sources are assumed to be the summation of the given mixture. \cite{liu2020separate} use a dual encoder-generator architecture with a cycle consistency loss to separate linear mixtures, more specifically the addition of two sources, using self-supervision. They explore the tasks of separating backgrounds and reflections, and separating albedo and shading. The Double-DIP method \citep{Gandelsman_2019_CVPR} seeks to solve image segmentation, image haze separation, and superimposed image separation with coupled deep-image-prior networks to predict a mask for separating two image layers. Separation techniques that use a single mask to separate two sources assume that there is a solution that can linearly separate a sample of mixed sources even though the process of generating said mask may be non-linear. \cite{Zou_2020_CVPR} propose their own method for separating superimposed images which uses a decomposition encoder with additional supervision from a separation critic and a discriminator to improve the quality of the source estimations. 
\cite{Kong_2019} also propose an adversarial approach for the same task on distorted MNIST digits such that the adversarial network is trained to distinguish real and generated sources. Recently, another work \citep{sunxiao2022} has also attempted source separation via adversarial networks with self-attention layers to separate superimposed multi-channel (RGB) images, but as before the adversarial network requires groundtruth source samples for training.

Biosignals processing tasks have benefited greatly from source separation approaches \citep{e24010055, 9826377, 9755920, Ong2019CASSCA, 9629984} and autoencoders have become an increasingly common strategy. Based on the work of \cite{Brakel2017LearningIF}, a recent work \citep{10285987} proposes using a shallow convolutional autoencoder network, but instead of using an adversarial network loss as the prior work has, their proposed approach applies three regularization functions on the predicted sources within the encoding space of the network. The ability of their proposed method to perform BSS on non-linear mixtures is limited by the single convolutional layer that maps the predicted sources to the mixture reconstruction. The authors apply their approach to the problem of BSS with high-density myoelectric signals. In a similar manner, \cite{MAYER2023105178} uses a shallow autoencoder with an orthogonality constraint to predict the discharge time of individual motor units from multichannel electromyogram recordings. The architecture can be thought of as a convolutional architecture in which the kernel size is one which may limit the effectiveness of this approach in applications where temporal or spatial dependencies should be modeled. Various works on BSS for ECG and PPG signals have shown promise in applications where the sources of interest may be assumed linearly mixed such as mixed fetal and maternal ECG signals \citep{RAMLI2020582}. Linear source separation techniques are also commonly used for the extraction and denoising of fetal ECG signals \citep{Cardoso1998MultidimensionalIC, Lathauwer1994FetalEE, articlesamenireza2010, 4042324, articlefatemi2017, articlejamshidantehrani2018}. In this paper, we explore an application of BSS to real-world ECG and PPG signals where these assumptions do not hold. See \autoref{appendix: related works} for more relevant works in the application of deep leanrning to source separation tasks.

\section{Proposed method}
\subsection{Preliminaries}
\paragraph{Multi-Encoder Autoencoders}\label{sec:mult-encoder autoencoders}
Autoencoders are a class of self-supervised neural networks that compress high-dimensional data into a low-dimensional latent space $z$ using an encoder network $E_{\theta}$, parameterized by $\theta$, and then reconstruct the input data from the latent space using a decoder network $D_{\phi}$, parameterized by $\phi$. During the training process, the encoder must learn to preserve the most salient information of the original data in order to make accurate reconstructions \citep{hinton1127647}.
\begin{gather} 
    z = E_{\theta}(x),\ \hat{x} = D_{\phi}(z) \\
    \min_{\theta, \phi} \mathcal{L}_{\text{recon.}}(x, D_{\phi}(E_{\theta}(x)))
\end{gather} 
Multi-encoder autoencoders \citep{ternes2022multiencoder} use a total of $N$ encoders $\bm{E_{\theta}} = \left[E^0_{\theta_{0}}\ E^1_{\theta_{1}}\ \ldots\ E^N_{\theta_{N}}\right]$, which all take the same input $x$ for this paper, unlike prior works, such that their separate encodings $\bm{z} = \left[z_{0}\ \ldots \ z_{N}\right]$ are concatenated along the channel axis before being passed into the decoder $D_{\phi}$ for reconstruction.
\begin{gather}
    \bm{z} = \bm{E_{\theta}}(x) = \left[E^0_{\theta_{0}}(x)\ E^1_{\theta_{1}}(x)\ \ldots\ E^N_{\theta_{N}}(x)\right] \\
    Z = E^0_{\theta_{0}}(x) \oplus E^1_{\theta_{1}}(x) \oplus \ldots\ E^N_{\theta_{N}}(x),\ \hat{x} = D_{\phi}(Z) \\
    \min_{\bm{\theta}, \phi} \mathcal{L}_{recon.}(x, D_{\phi}(\bm{E_{\theta}}(x)))
\end{gather}
\paragraph{Blind Source Separation}

There are three common mixing scenarios considered in BSS literature: determined mixtures, overdetermined mixtures, and underdetermined mixtures \citep{bookcommonpierre2010, inproceedingsdevileyannick, inproceedingsduarteleondardo}. For determined mixtures, the number of sources is equal to the number of observed mixtures. In the overdetermined scenario, the number of observed mixtures is greater than the number of sources. Finally, in the case of underdetermined mixtures, the number of sources is greater than the number of observed mixtures, and in this paper, single-channel underdetermined mixtures are explored. By definition, single-channel mixtures with more than one source are underdetermined. However, it should be noted that the proposed method is not explicitly limited to this scenario. Undetermined mixtures are particularly challenging because without additional priors the possible solutions may be infinite \citep{HYVARINEN1999429}.

A simplified formal definition of blind source separation is given as follows: For some set of $N$ source signals $\bm{s}(t)$ and some set of $K$ noise signals $\bm{n}(t)$, a mixing system $\mathcal{F}$ operates on $\bm{s}(t)$ and $\bm{n}(t)$ to produce a mixed signal $x(t)$. Further, the mixing system $\mathcal{F}$ may be non-linear and non-stationary in time or space $t$.
\begin{gather} 
    \bm{s}(t) = \left[s_{0}(t)\ s_{1}(t)\ \ldots\ s_{N}(t)\right] \\
    \bm{n}(t) = \left[n_{0}(t)\ n_{1}(t)\ \ldots\ n_{K}(t)\right] \\
    x(t) = \mathcal{F}(\bm{s}(t), \bm{n}(t), t)
\end{gather}
In BSS, the goal is either to find an inverse mapping of the mixing system $\mathcal{F}^{-1}$ or extract the source signals $\bm{s}(t)$ given only the mixture $x$. Though modeling with linear systems may be sufficient in many cases, the problem of BSS with non-linear mixtures can be especially challenging due to the additional ambiguity it creates in the mapped feature space \citep{NIPS2001_cf2226dd, inbookdeville, articlesimasfilho2012}. The degree of ambiguity created by the mixing system directly relates to the important issue of identifiability or separability in BSS literature. More information about this topic and its relevance to the proposed method can be found in \autoref{appendix: Indentifiablility and Separability with Non-Linear Mixtures}.

\subsection{Blind Source Separation with Multi-Encoder Autoencoders}
In this paper, we train a multi-encoder autoencoder for reconstructing underdetermined single-channel input mixtures where the encoders specialize in the feature subspaces of different sources as an emergent property of learning from the feature distribution of the data. Specifically, the encoders unmix the sources within the encoding space of the network, and the decoder remixes the sources such that each source component removed in the encoding space results in the source being removed from the output reconstruction. In \autoref{fig: Source Separation Training}, the training phase is shown during which all encoders $\bm{E}$ are active and the expected output of the network is a reconstruction $\hat{x}$ of the input signal $x$. However during inference, shown in \autoref{fig: Source Separation Inference}, only one encoder is active and the expected output is the source signal $s^n$ that corresponds to the source feature subspace that the active encoder $E^{n}$ specializes in. In place of the encodings from non-active encoders, zero matrices of the same size are concatenated to the active encoder's output encoding, keeping the position of the active encoding in place.
\paragraph{Multi-Encoder Design and Regularization}
\label{section: multi-encoder design and regularization}
As stated in the description of multi-encoder autoencoders (i.e., \autoref{sec:mult-encoder autoencoders}), $N$ encoders $\bm{E_{\theta}} = \left[E^0_{\theta_{0}}\ E^1_{\theta_{1}}\ \ldots\ E^N_{\theta_{N}}\right]$, parameterized by $\bm{\theta} = \left[\theta_{0}\ \theta_{1}\ \ldots\ \theta_{N}\right]$, are chosen and produce encodings $\bm{z} = \left[z^{0}\ z^{1}\ \ldots \ z^{N}\right]$ for input variables $x \in \mathbb{R}^{C \times M}$ where $M$ represents an arbitrary number of dimensions. Prior knowledge of the number of sources is not strictly necessary for choosing the number of encoders as overestimating can result in \emph{dead encoders} that have no contribution to the final source reconstruction though they may still contribute information about the mixing system. An occurrence of such a dead encoder is exemplified later in \autoref{sub: tri circ}. 

Batch normalization \citep{pmlr-v37-ioffe15} is applied to all layers with the exception of the output of each encoder. We apply an L2 regularization loss \eqref{eq: z loss} to the output of each encoder during training as follows to prevent these values from growing too large and causing exploding gradients.
\begin{gather} 
\label{eq: z loss}
    \mathcal{L}_{\text{z}} = \lambda_{\bm{z}} \frac{1}{N h} \sum_{z^n \in \bm{z}} \|z^n\|_2^2
\end{gather} 
where $\|z^n\|_2^2$ represents the Euclidean norm of some encoding $z^n$ squared, the coefficient $\lambda_{\bm{z}}$ controls the penalty strength, and $h$ represents the size of the encoding (product of the channel size and spatial size). 
\begin{gather} 
\label{eq: concat}
    Z = E^{0}(x) \oplus E^{1}(x) \oplus \ldots \oplus E^{N}(x)
\end{gather} 
where the $\oplus$ operator represents concatenation along the channel dimension. As the final step of the encoding phase, the encodings $\bm{z}$ are concatenated along the channel dimension to produce the input for the decoder $Z$. One final important consideration for designing the encoders is choosing a proper encoding size. The size should not be so small as to prevent good reconstructions but also not so large that every encoder generalizes to the entire feature space rather than specializing.
\paragraph{Decoder Design and Regularization}
The proposed method uses a standard single decoder architecture $D_{\phi}$ parameterized by $\phi$, as described previously in \autoref{sec:mult-encoder autoencoders}, with input $Z$ such that the reconstruction $\hat{x} = D_{\phi}(Z)$. In this paper, two novel regularization techniques are designed for guiding the network toward conducting blind source separation. See \autoref{appendix: regularization} for more details about regularization for the proposed method.
\begin{figure}
  \centering
  \begin{subfigure}[]{0.325\textwidth}
        \centering
            \includegraphics[width=1.0\linewidth]{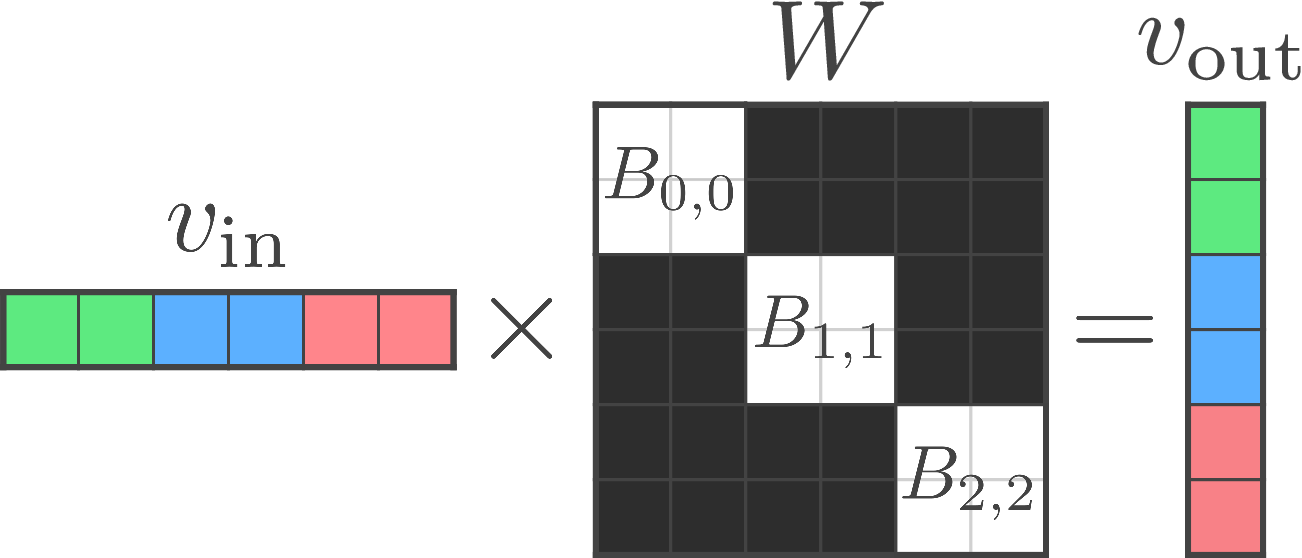}
        \caption{}
        \label{fig: unmixed sep}
  \end{subfigure}\hspace{0.1\textwidth}%
  \begin{subfigure}[]{0.325\textwidth}
        \centering
        \includegraphics[width=1.0\linewidth]{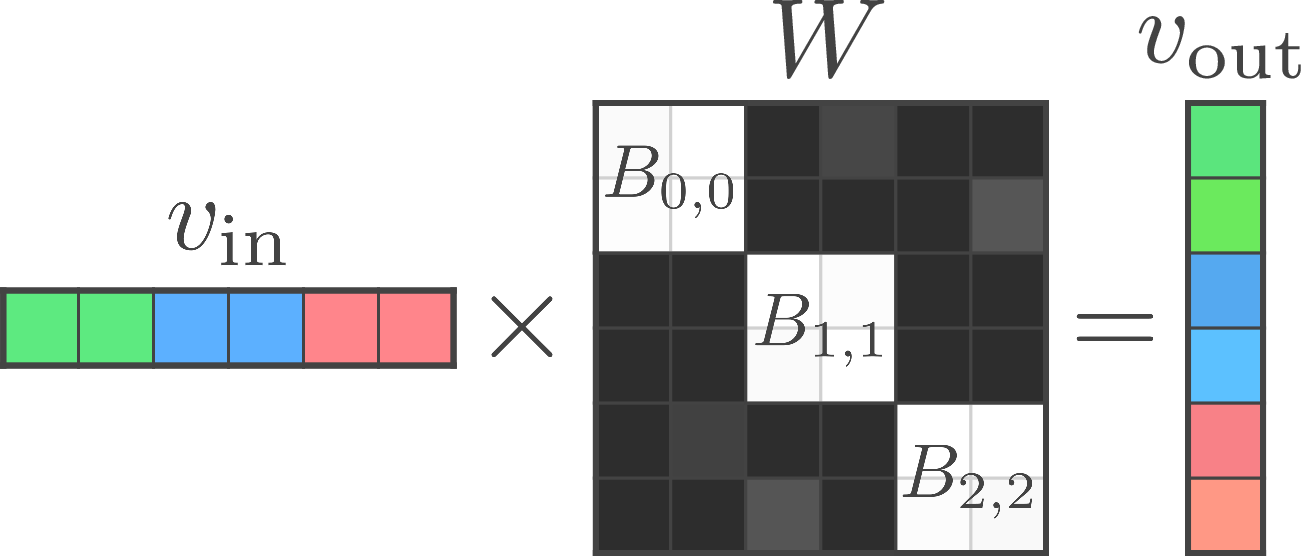}
        \caption{}
        \label{fig: mixed sep}
  \end{subfigure}
  \caption{Illustration of the effect that the \emph{sparse mixing loss} has on weight matrices while decoding separate encoding spaces.}
  \label{fig: separation loss}
\end{figure}

A \textbf{sparse mixing loss} is proposed for each layer in the decoder, with the exception of the output layer, to encourage sparse mixing between source encodings and their mappings throughout the decoder. For a given layer in the decoder the weights are first segmented into a set of blocks or non-overlapping submatrices $\bm{B}$ where the number of blocks is equal to the number of encoders squared i.e. $N^2$. For each block $B_{i,j}$ in the weight $W$, with elements $w \in W$, the height and width are equal to the input size $C_{\text{in}}$ and the output size $C_{\text{out}}$ of $W$ divided by the number of encoders $N$. For convolutional layers, this process is generalized with respect to the channel dimensions. Then all submatrices $B_{i, j}$ in the set of submatrices $\bm{B}$, where $i \neq j$ i.e. off-diagonal blocks, are decayed towards zero using the L1 norm.
\begin{gather} 
\label{eq: mixing loss}
W = \bm{B} = \begin{bmatrix}
    B_{0,0} & B_{0,1} & \cdots & B_{0,N}\\
    B_{1,0} & B_{1,1} & \cdots & B_{1,N} \\
    \vdots  & \vdots  & \ddots & \vdots \\
    B_{N,0} & B_{N,1} & \cdots & B_{N,N}\\
\end{bmatrix},\ B_{i,j} = \begin{bmatrix}
    w_{\frac{C_{\text{in}}}{N} i,\frac{C_{\text{out}}}{N} j} & \cdots & w_{\frac{C_{\text{in}}}{N}i,\frac{C_{\text{out}}}{N} (j+1)}\\
    \vdots  & \ddots & \vdots \\
    w_{\frac{C_{\text{in}}}{N} (i+1),\frac{C_{\text{out}}}{N} j} & \cdots & w_{\frac{C_{\text{in}}}{N} (i+1),\frac{C_{\text{out}}}{N} (j+1)}\\
\end{bmatrix} \\
\mathcal{L}_{\text{mixing}} = \sum_{B_{i,j} \in \bm{B}}^{i \neq j} \alpha \|B_{i,j}\|_1 \\
\alpha = \frac{1}{\frac{C_{\text{in}}}{N} \frac{ C_{\text{out} }}{N}},\ \text{or} \label{eq: alpha 1} \\
\alpha_{i,j} = \begin{cases} \label{eq: alpha 2}
    \ \frac{1}{(N - i) \frac{C_{\text{in}}}{N} \frac{ C_{\text{out} }}{N}},       & \quad \textrm{if } j > i\\
    \ \frac{1}{(N - (N - i)) \frac{C_{\text{in}}}{N} \frac{ C_{\text{out} }}{N}}, & \quad \textrm{if } j < i
  \end{cases}
\end{gather} 
where $\|B_{i,j}\|_1$ denotes the L1 norm of some block $B_{i,j}$. The sparse mixing loss is not applied to the final output layer of the network. In our standard formulation, the scalar $\alpha$ is an additional term equal to the reciprocal of the number of elements in each block $B_{i,j}$. However, for our experiments with PPG and ECG signals, another scaling scheme that assigns different penalties depending on each block's position is devised in consideration of the linear and non-linear relationships between different sources in PPG or ECG signals, as is seen in \autoref{eq: alpha 2}. The role of the proposed \emph{sparse mixing loss} is illustrated in \autoref{fig: separation loss}. For the weight matrix $W$, the smallest squares have a value of one if they are white and a value of zero if they are black. In the input and output vectors, $v_{\text{in}}$ and $v_{\text{out}}$, the red, green, and blue colors represent three separate source encodings or mappings. In the case that all off-diagonal weights are exactly zero as illustrated by \autoref{fig: unmixed sep}, no mixing occurs between the source encodings until the output layer (a convolutional layer with a kernel size of 1) of the network meaning that the signals are linearly mixed with the possibility of one non-linearity at the output\footnote{For our experiments a sigmoid function is applied to the output for training with the binary cross entropy loss. If the triangles \& circles mixtures did not include a distortion kernel after the sigmoid, the described scenario would be a valid solution.}. However, for non-linearly mixed signals, off-diagonal blocks in each layer are expected to have at least some non-zero values as illustrated in \autoref{fig: mixed sep}.

During source inference, all encodings except the encoding associated with the source of interest are masked out with zero arrays. However, even when the encodings are masked, the biases and weights that do not decode the target source may still contribute to the final output causing undesirable artifacts. Thus, a \textbf{zero reconstruction loss} is proposed to ensure that masked source encodings have minimal contribution to the final source estimation. For the zero reconstruction loss, an all-zero encoding vector $Z_{\text{zero}}$ is passed into the decoder, and the loss between the reconstruction $\hat{x}_{\text{zero}}$ and the target $x_{\text{zero}}$, an all-zero vector equal to the output size, is minimized. This encourages the network to adjust the weights and bias terms of each layer in a way that encourages inactive encoding spaces to have little or no contribution to the output prediction. This loss is implemented as a second forward pass through the network, but the gradients are applied in the same optimization step as the other losses. While conducting this second forward pass for the decoder, the affine parameters of each normalization layer are frozen i.e. no gradient is calculated for the affine parameters when calculating the zero reconstruction loss at each step.
\begin{gather} 
    \min_{\phi} \mathcal{L}_{\text{zero recon.}}(x_{\text{zero}}, D_{\phi}(Z_{\text{zero}}))
\end{gather} 
As for the normalization layers in the decoder, group normalization \citep{Wu_2018_ECCV} is used such that the number of groups matches the number of encoders. Group normalization is applied to all layers in the decoder except the output layer. Group normalization is ideal for the proposed method as it can independently normalize each encoding mapping throughout the decoder. Additional notes about the addition of alternative losses can be found in \autoref{appendix: regularization}.

\paragraph{Training and Inference}
The entire training process can be summarized as minimizing the reconstruction loss between the input and output of the multi-encoder-single-decoder autoencoder model with additional regularization losses that help ensure quality source separation during inference: encoding regularization $\mathcal{L}_{\text{z}}$, sparse mixing $\mathcal{L}_{\text{mixing}}$, and zero reconstruction loss $\mathcal{L}_{\text{zero recon.}}$. The training process is also summarized in \autoref{algo: train} within the appendices. The final training loss is expressed with the formula below.
\begin{gather} 
    \mathcal{L}_{\text{total}} = \mathcal{L}_{\text{recon.}} + \mathcal{L}_{\text{mixing}} + \mathcal{L}_{\text{zero recon.}} + \mathcal{L}_{\text{z}}
\end{gather} 
In this paper, only min-max scaling on the input $x$ and a binary cross-entropy loss for the reconstruction terms $\mathcal{L}_{\text{recon.}}$ and $\mathcal{L}_{\text{zero recon.}}$ are considered.

During inference, source separation is performed by leaving the $n^{\text{th}}$ encoder $E^{n}_{\theta_{n}}$ active and masking out all other encodings with zero vectors $\bm{0}$. The concatenation of the active encoding with the masked encodings $Z^n$ are passed into the decoder to give the source estimation $\hat{s}^n$ as illustrated in \autoref{fig: Source Separation Inference} and described in \autoref{algo: inference} located in the appendices.
\begin{gather} 
    Z^n = \bm{0} \oplus \ldots \oplus E^{n}_{\theta_{n}}(x)  \oplus \ldots \oplus \bm{0} \\
    \hat{s}^n = D_{\phi}(Z^n)
\end{gather} 
For all encoders, this step is repeated to get each source prediction $\hat{s}^n$ for some input $x$. Cropping the edges of the predicted source when using convolutional architectures may be necessary due to information loss at the edges of convolutional maps which more significantly impacts source signals that have only a minor contribution to the mixed signal $x$. For the circle and triangles dataset there is no need for cropping source estimations because the two sources have equal contributions to the mixture in terms of amplitude, however, this step is necessary for the ECG and PPG experiments.

\section{Experimental Evaluation}
In this section, we evaluate the effectiveness of our proposed methodology in comparison to existing methods. In Sections \ref{sub: tri circ} and \ref{sub: ppg ecg}, we conduct experiments with the triangles \& circles toy dataset and on ECG and PPG data for respiratory source extraction, respectively. 
\subsection{Triangles \& Circles Toy Dataset} \label{sub: tri circ}
A toy dataset comprised of PNL mixtures of triangle and circle shapes is generated for an intuitive analysis of the proposed method. The mixing system used in this dataset has visually easy-to-understand dynamics and serves as a simple introduction to the behavior of the proposed method in practice. Our approach to generating a synthetic dataset for BSS is distinct from similar works \citep{liu2020separate} as it uses non-linear mixing and convolves the PNL mixture with a distortion kernel designed to intentionally create minor spatial indeterminacies in the final mixture.

For this experiment, a fully convolutional multi-encoder autoencoder, detailed in \autoref{fig: tri circl model spec} (located in \autoref{appendix: model details}), is trained on the introduced toy dataset using the proposed method. \emph{Though there are only two sources for this dataset, three encoders are implemented to show the emergence of a dead encoder.} Hyperparameter settings for the optimizer and all loss terms can be found in \autoref{appendix: Optimizer and Hyperparameter Settings}.

\paragraph{Data Specifications} The Python Imaging Library \citep{umesh2012image} is used to generate the training and validation sets of post non-linearly mixed single-channel triangles and circles convolved by a distortion filter. $150,000$ triangle and circle image pairs, $x_{\triangle}$ and $x_{\bigcirc}$, are generated such that they both have uniform random positions within the bounds of the final image and uniform random scale between 40\% and 60\% the width of the image. Where the triangle and circle shapes exist within the image the luminance value is 100\% or a value of $1.0$, and everywhere else within the image is a luminance of 0\% or a value of $0.0$ (after resampling, the values at the edges of shapes may vary). The sources are generated with a height and width of $128$ pixels and then downsampled to a height and width of $64$ pixels using bilinear resampling. Then for each pair of triangle and circle images, the following PNL mixing system is applied.
\begin{gather} \label{eq: tri circ mxing}
    x_{\text{mixed}} = \text{{Scaled}}(\sigma(\frac{\beta}{2}(x_{\triangle} + x_{\bigcirc}))) \ast \text{Distort. Kernel} \\
    \text{{Scaled}}(x) = \frac{{x - \text{{min}}(x)}}{{\text{{max}}(x) - \text{{min}}(x)}}
\end{gather} 
where $\sigma$ refers to the sigmoid function, $\text{Scaled}$ refers to a min-max scaling function, and $\ast$ refers to the convolution operation. The distortion kernel is designed such that the edges of the combined shapes become visually distorted in a shifted and blurred manner. In addition, the distortion kernel is horizontally flipped with a $50\%$ probability. The exact kernel used and additional details can be found in \autoref{appendix: selection of distortion kernel}. \emph{The $\beta$ argument controls the degree to which the luminance of non-overlapping regions matches overlapping regions.} In the final triangles \& circles dataset experiment, $\beta$ equals 6. To create the training and test splits, 80\% of the generated data is reserved for training and the remaining 20\% of the data becomes the test set. 

\begin{figure}
  \centering
  \begin{subfigure}[]{0.35\textwidth}
        \centering
            \includegraphics[width=1.0\linewidth]{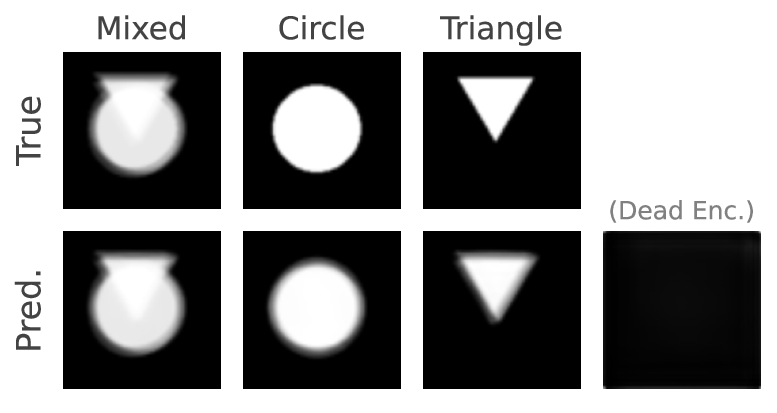}
        \caption{}
        \label{fig: tri circ unmixed sep}
  \end{subfigure}\hspace{0.1\textwidth}%
  \begin{subfigure}[]{0.35\textwidth}
        \centering
        \includegraphics[width=1.0\linewidth]{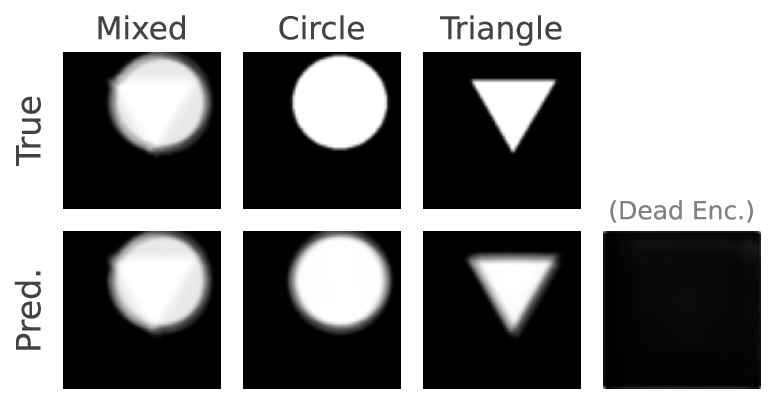}
        \caption{}
        \label{fig: tri circ mixed sep}
  \end{subfigure}
  \caption{Sample predictions from the triangles \& circles dataset experiment using the proposed method.}
  \label{fig: tri circ}
\end{figure}
\begin{table}
\caption{Average MSE and SSIM scores on the test set for triangle sources and circle sources. A supervised AE architecture trained to directly predict the sources is included for comparison.}
\centering
    \begin{tabular}{lllll}
        \toprule
        Method & Triangle MSE $\downarrow$ & Circle MSE $\downarrow$ & Triangle SSIM $\uparrow$ & Circle SSIM $\uparrow$\\
        \midrule
        Ours (BSS) & 0.00636 & 0.00478 & 0.89860 & 0.86914\\
        AE (Supervised) & 0.00005 & 0.00007 & 0.99890 & 0.99873\\
        \bottomrule
    \end{tabular}
\label{table: tri_circ}
\end{table}
\paragraph{Evaluation and Analysis} \autoref{fig: tri circ} shows two samples with the predicted reconstruction and source estimations from the trained model along with the ground truth mixture and sources (a video showing the training progression can be found in the supplementary material). To evaluate the ability of the proposed method to reconstruct sources, the mean squared error (MSE) and structural similarity index measure (SSIM) \citep{2004ITIP...13..600W} between source predictions and ground truth sources is calculated and averaged over the entire test for each metric set as shown in \autoref{table: tri_circ}. MSE and SSIM are common metrics used in computer vision for evaluating the similarity of two images. Typical metrics used in BSS literature such as the correlation coefficient are not suitable for this toy problem due to the possibility of rows and columns with zero variance in each image. SSIM creates a score between $-1$ and $1$ where negative one represents complete dissimilarity, zero represents no similarity, and one represents complete similarity. Along with the results of the proposed method, the results of a deep convolutional autoencoder (labeled as AE) trained to directly predict the sources in a supervised manner are provided for comparison. See \autoref{appendix: model details} for details on the architecture used. The autoencoder results show that it is possible to accurately predict the sources given that the exact sources are known. In the unsupervised BSS case, the non-linearity and spatial indeterminacies introduced by the mixing system result in an infinite number of solutions being possible, but because the proposed approach is able to learn structural information about the sources from the distribution of the dataset (allowing the multiple encoders to specialize in representing a source), an approximate and consistent solution can be found even though the mixing system varies from sample to sample. For each mixture in the dataset, there is an exact (non-convolutive) linear mapping that will produce the true sources, but there is no linear solution that generalizes to all samples. Thus, because the network used is fully convolutional and does not have access to the sources, it must learn the dynamics of the mixing system from the dataset distribution rather than memorizing or predicting unique linear solutions for every possible mixture. We show further that the model generalizes to the mixing system using two additional experiments in \autoref{appendix: generalizing to the mixing system}. For further notes about the applicability of other common BSS separation techniques for this problem, see \autoref{appendix: Application of Common BSS Approaches}.
\begin{figure}
  \centering
  \includegraphics[width=0.75\textwidth]{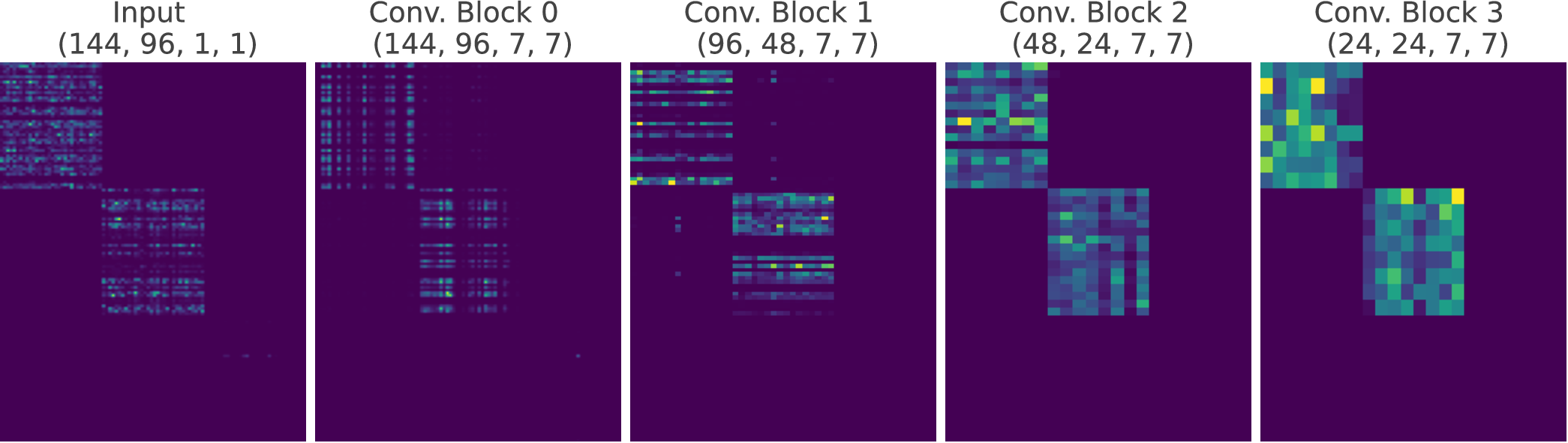}
  \caption{Visualization of weights in the decoder after training a model on the triangles \& circles dataset with the proposed method.}
  \label{fig: tri_circ_weights}
\end{figure}

In \autoref{fig: tri_circ_weights}, the absolute values of the weights for each layer in the decoder are summed along the spatial dimensions, and the decoder input layer is transposed. The application of the sparse mixing loss is made apparent by the many values close to zero in the off-diagonal blocks except for a few weights responsible for mixing within the on-diagonal blocks (visible in \emph{Conv. Block 1}). These shared connections establish the relationship between the two source encoding spaces within the decoding or \emph{remixing} process. The degree of sparsity in the off-diagonal blocks reflects the simplicity of the mixing system used. The bottom right on-diagonal block is also noted to have values almost completely at zero throughout the decoder indicating a \emph{dead encoding space} as described in \autoref{section: multi-encoder design and regularization}.

\subsection{ECG and PPG Respiratory Source Separation} \label{sub: ppg ecg}
Two practical experiments using real-world single-channel biosignal recordings from a polysomnography (PSG) study are conducted using the proposed method. Specifically, the blind source separation of respiratory source signals from unprocessed ECG and PPG signals, which are typically used for the analyses of cardiovascular health, are considered. There is a vast literature on extracting respiratory rate or respiratory signals from ECG and PPG using various heuristic, analytical, or source separation techniques \citep{soni, Charlton_2016, VANGENT2019368, iet:/content/conferences/10.1049/cp.2015.1654, 9629984, DAVIES2023104992, 8856301, Lipsitz1995HeartRA, varonarticle, s20113238}. However, BSS methods applied to respiratory signal extraction from ECG or PPG are challenging because the cardiovascular, respiratory, and other sources (such as movement) may not be assumed entirely linearly mixed and may be correlated in time \citep{Jamek2004NonlinearCI, Kanters1997-vt, articleplastiamirjana2023}. In addition, ECG or PPG signals can be influenced by respiratory-induced amplitude modulation, frequency modulation, or baseline wander at different times in a signal depending on a number of physiological factors such as respiratory sinus arrhythmia \citep{Hirsch1981RespiratorySA}, or the various mechanical effects of respiration on the position and orientation of the heart \citep{surawicz2008chou}. In terms of BSS, this variation translates to a mixing system that changes in time which adds to the difficulty of the problem. For these reasons, many BSS methods fail to separate respiratory sources in a meaningful way, and thus heuristics are often relied upon \citep{iet:/content/conferences/10.1049/cp.2015.1654} even if they involve BSS techniques \citep{6144719, s20113238}. See \autoref{appendix: efficacy of common bss approaches} for further discussion on the BSS approaches tested.

For both the ECG and PPG experiments, the same fully convolutional multi-encoder autoencoder, detailed in \autoref{fig: ecg ppg model spec} (located in the appendices), is trained using the proposed method. Additionally, all hyperparameter settings are located in \autoref{appendix: Optimizer and Hyperparameter Settings ecg_ppg}.

\paragraph{Data Specifications} The Multi-Ethnic Study of Atherosclerosis (MESA)  \citep{zhang2018national, chen2015racial} is a research project funded by the NHLBI that involves six collaborative university clinics. The MESA Sleep study provides PSG data which includes simultaneously measured ECG, PPG, and other respiratory-related signals. PPG signals capture changes in blood volume and are used for measuring heart rate, oxygen saturation, and other cardiovascular parameters. ECG signals record the heart's electrical activity and are used to assess cardiac rhythm, detect abnormalities, and diagnose cardiac conditions. Both PPG and ECG signals play important roles in monitoring and diagnosing cardiovascular health. 

From the PSG data, unprocessed PPG and ECG signals are extracted for the purpose of separating a respiratory source using the proposed method which is done without the supervision of a reference respiratory source or the use of any heuristics or strong priors. Thoracic excursion recordings, which measure the expansion of the chest cavity, and nasal pressure recordings, which measure the change in airflow through the nasal passage, are extracted as reference respiratory signals for verifying the quality of the predicted respiratory source separations. For this work, $1,000$ recordings of the total $2,056$ PSG recordings are randomly selected from the MESA PSG dataset. Then, after randomly shuffling the recordings, the first $500$ recordings of this subset are used for training and the remainder are used for validation. As a first step, the PPG and ECG recordings for each PSG recording are resampled from the original 256hz to 200hz. Then each recording is divided into segments that are $12,288$ samples in length, or approximately $60$ seconds. As a final step, the segments are scaled using min-max scaling. No other processing or filtering is applied to the segments, however, some segments are removed if a segment's average heart rate is simply detected to be outside the range of $40$-$180$ beats per minute. For evaluation of the extracted respiratory sources, the NeuroKit2 open source library \citep{Makowski2021neurokit} is used to process the respiratory signals and then extract the respiratory rate.
\begin{figure}
  \centering
  \begin{subfigure}[]{1.0\linewidth}
        \centering
        \includegraphics[width=0.725\textwidth]{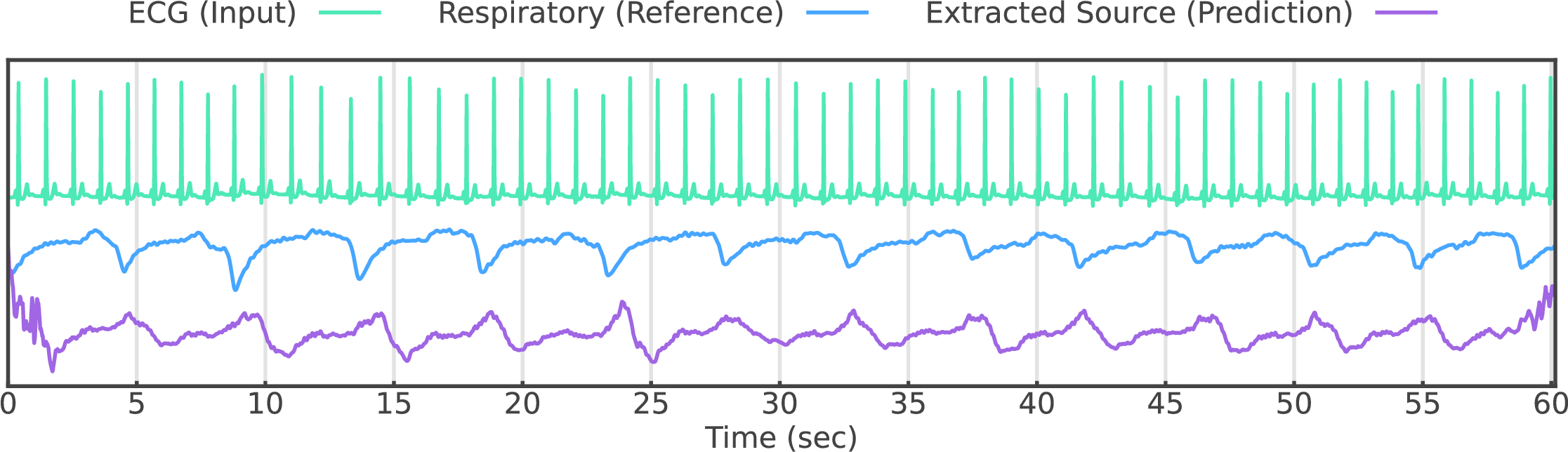}
        \caption{ECG respiration extraction via proposed BSS method.}
        \label{fig: ecg}
  \end{subfigure}
  \begin{subfigure}[]{1.0\linewidth}
        \centering
        \includegraphics[width=0.725\textwidth]{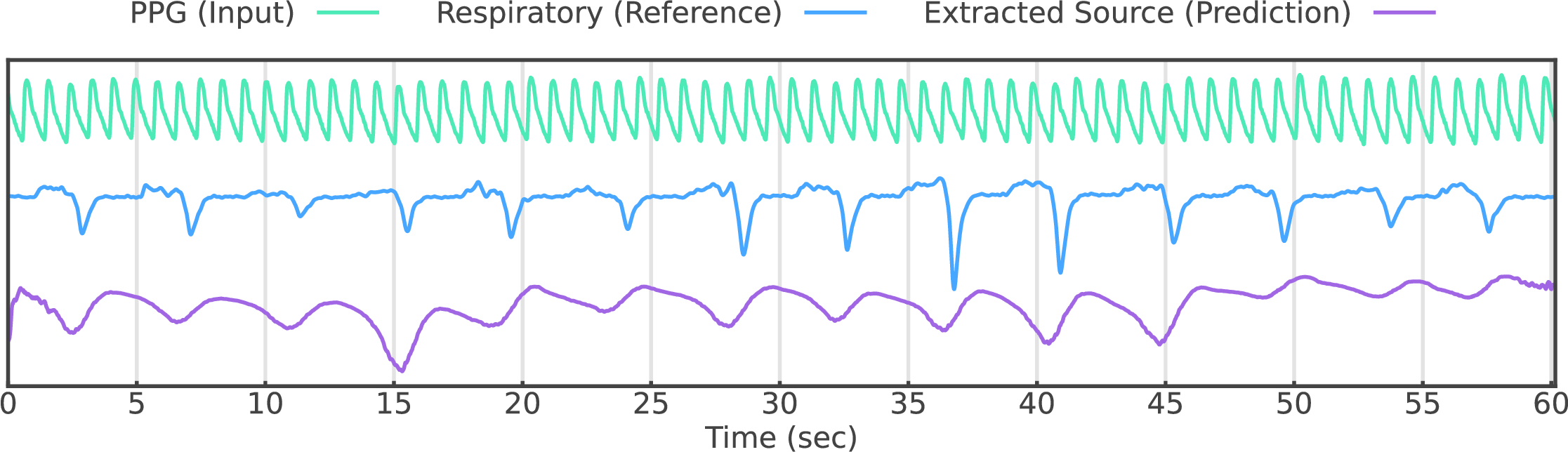}
        \caption{PPG respiration extraction via proposed BSS method.}
        \label{fig: ppg}
  \end{subfigure}
  \caption{Sample graphs showing our extracted respiration sources from PPG and ECG along with a simultaneously measured reference respiratory signal.}
  \label{fig: ecg ppg preds}
\end{figure}
\paragraph{Evaluation and Analysis} In \autoref{fig: ecg ppg preds}, a simultaneously measured reference respiration signal, and the extracted respiration source signal using our method are visualized for both ECG and PPG experiments (more samples are in the supplementary material). In \autoref{fig: ecg} the reference respiration signal shown is thoracic excursion, and for \autoref{fig: ppg} the reference respiratory signal is nasal pressure. While the predicted respiratory sources are the result of the respiratory system's functions they may also represent subtly different physiological processes than those measured by nasal flow or thoracic excursion (it may be more appropriate to describe these two signals as pseudo-references in this scenario so as not to confuse the common meaning of "reference" in source separation literature). Thus the predicted respiratory source and the reference respiration signal may have subtle or obvious differences in feature expression. For example, the predicted respiration source may have unique amplitude modulation patterns or appear flipped relative to the respiration reference as seen in \autoref{fig: ecg}. For these reasons, we cannot effectively evaluate the proposed method on this challenging problem with more standard metrics used in the source separation literature such as SAR, SIR, or SDR \citep{1643671vincent} (which are commonly used metrics for audio source separation tasks) because the ground truth respiratory source within a real-world ECG or PPG signal is not knowable and therefore what constitutes interference or artifacts is also unknowable. Even the reference signals have considerably different feature expressions from one another (as seen in the supplementary material). Instead, the proposed method is evaluated by extracting respiratory rate from the predicted source signals and for the two reference respiratory signals, nasal flow and thoracic excursion, which is achieved by measuring the periods between inhalation onsets or troughs for one-minute segments. Then the mean absolute error (MAE) between the respiratory rates the source estimation and each of the reference signals is calculated to determine the degree to which the predicted respiration sources preserve respiratory information relative to these references.
\begin{table}
  \caption{The test mean absolute error (MAE) in breaths per minute between each method of respiratory signal extraction (BSS methods, heuristic approaches, and two additional self-supervised baselines), and the two reference respiratory signals.}
  \label{ecg-ppg-table}
  \centering
    \begin{tabular}{cccccc}
        \toprule
        Method (Type) & \multicolumn{2}{c}{Breaths/Min. MAE $\downarrow$}
        & Method (Type) & \multicolumn{2}{c}{Breaths/Min. MAE $\downarrow$}\\\cmidrule(r){1-3}\cmidrule(r){4-6}
        BSS & Nasal Press. & Thor. & Heuristic & Nasal Press. & Thor.\\\cmidrule(r){1-3}\cmidrule(r){4-6}
        Ours (PPG) & 1.51 & 1.50 & \citep{soni} (ECG) & 2.38 & 2.18\\
        Ours (ECG) & 1.73 & 1.59 & \citep{Charlton_2016} (ECG) & 2.38 & 2.17\\
        \emph{See \autoref{appendix: ecg ppg} for} & & & \citep{VANGENT2019368} (ECG) & 2.27 & 2.05 \\
        \emph{failed results.} & & & \citep{iet:/content/conferences/10.1049/cp.2015.1654} (ECG) & 2.26 & 2.07\\
        & & & \citep{Lindberg1992MonitoringOR} (PPG) &  2.38 & 2.02 \\
        & & & \citep{Lindberg1992MonitoringOR} (ECG) & 2.82 & 2.42 \\
        \midrule
        Supervised \\(Nasal Press. as Target) & & & Direct Comparison & &\\
        \cmidrule(r){1-3}\cmidrule(r){4-6}
        AE (PPG) & 0.46 (0.87) & 2.07 (1.71) & Thor. & 1.33 & --\\
        AE (ECG) & 0.48 (1.03) & 2.16 (1.69)\\
        \bottomrule
    \end{tabular}
\end{table}

The experimental results in \autoref{ecg-ppg-table} show the test mean absolute error (MAE) in breaths per minute between each method of respiratory signal extraction and the two reference respiratory signals, nasal pressure and thoracic excursion (abbreviated as \emph{nasal press.} and \emph{thor.}) Here, \emph{type} refers to the input signal that the method extracts the respiratory signal from, either ECG or PPG. The primary result shows that the proposed method for BSS performs better than the competing heuristic approaches designed specifically for the task of respiratory signal extraction. Two additional baselines are included as a point of reference: the results of a supervised model trained to predict the nasal pressure signal from the two types of input signals (similar to other works \citep{9629984, DAVIES2023104992, 8856301}) and a direct comparison between the thoracic excursion signal and nasal flow signal. The supervised approach, employing a standard autoencoder (AE) architecture, generally performs well at predicting the reference respiratory signal it is trained to reconstruct, nasal pressure, but generalizes poorly to the other reference respiratory signal, thoracic excursion. Early-stopping does result in better generalization (shown in parenthesis next to the primary result). See \autoref{appendix: model details} for more details about the architecture used for the supervised autoencoder approach.

\subsection{Code Availability} \label{sub: code}
Our implementation of the experiments described in this paper is available on GitHub at the following link: \href{https://github.com/webstah/self-supervised-bss-via-multi-encoder-ae}{\url{github.com/webstah/self-supervised-bss-via-multi-encoder-ae}}.
\section{Conclusion}
The method for BSS using multi-encoder autoencoders introduced in this paper presented a novel data-driven solution to BSS without strong application-specific priors or the assumption that sources are linearly separable from the given mixture. The ability of the multiple encoders to discover the feature subspaces of sources and of the decoder to reconstruct those sources was demonstrated on two datasets. The triangles \& circles dataset served as an intuitive example of how the proposed method can learn from the distribution of the dataset to perform BSS even when the distinguishing features and distribution of the sources are a limiting factor. Then to demonstrate a real-world application of the proposed method, an experiment for extracting respiratory signals from ECG and PPG signals in a self-supervised and end-to-end manner with limited data-cleaning was devised. The method proposed in this paper and similar deep learning approaches to source separation present an exciting research path within the topics of BSS and self-supervised learning.
\subsection{Broader Impact Statement}
While this work shows that meaningful source separation occurs with the ECG and PPG experiments, the ability of the proposed method, or any BSS approach, to reconstruct exact sources should not be overstated and this fact must be taken into consideration before applying the proposed method in any real-world context especially those involving medical diagnosis or decisions that may negatively impact any individuals, demographics, or society as a whole. Further investigation is necessary to understand the shortcomings of the proposed method despite the positive results presented in this paper.
\subsection{Acknowledgements}
We want to express our sincere thanks to Dr. Masoud R. Hamedani and Dr. M. Dujon Johnson for their help in editing and providing valuable suggestions to enhance the clarity and flow of this paper. Additionally, we thank Dr. Jeremy Wurbs for their engaging discussions that helped kick-start our research journey and Donghyeon Lee for helping improve the mathematical notation used in this paper.

This work was supported by the Technology Development Program (S3201499) funded by the Ministry of SMEs and Startups (MSS, Korea).
\bibliographystyle{acm}
\bibliography{references} 
\newpage
\appendix
\appendixpage
\section{Additonal Related Works} \label{appendix: related works}
Here we provide an overview of additional related works, that while not all directly related to the focus of this paper, add insight for the reader into the current state of deep learning for source separation and the approaches that have been proposed in other domains of interest.

In the audio domain, source separation is applied to various fields; for example, in speech separation where the voices of different individuals are the source signals, and also in music source separation, where the individual instruments that comprise the mixed audio are the target sources. 
In the literature, different methods for source separation based on deep learning algorithms have been proposed. Wave-U-Net \citep{DBLP:conf/ismir/StollerED18} brings end-to-end source separation using time-domain waveforms. The deep attractor network (DANet), \citep{Chen2016DeepAN} uses an LSTM network to map mixtures in the time-frequency domain into an embedding space which is then used with so-called attractors, taking inspiration from cognitive speech perception theories in neuroscience, to create masks for each source.

Generating masks on a per-source basis makes the assumption that the source is linearly separable from the mixture which is a reasonable and often necessary assumption to make for many problems. Deep clustering \citep{7471631} starts by embedding time-frequency representations of a mixed audio signal, then using K-means clustering to segment multiple sources, and lastly with the informed segmentation uses sparse non-negative matrix factorization to perform source separation. TaSNet \citep{Luo2017TaSNetTA} uses a three-stage network with an encoder that estimates a mixture weight, a separation module that estimates source masks, and a decoder for reconstructing source signals. Other works make extensions to TaSNet such as the dual-path RNN \citep{9054266} that improves performance while reducing the model size, Conv-TasNet \citep{luo2019} which brings a fully convolutional approach to the prior method, or Meta-TaSNet \citep{Samuel2020MetaLearningEF} which introduces meta-learning to generate the weights for source extraction models. \cite{Brakel2017LearningIF} employ an adversarial network to enforce independence between unmixed sources within the encoding space of a feedforward autoencoder architecture by minimizing the adversarial loss between source predictions of input mixtures and resampled sources whereby the individual sources produced by one mixture are randomly combined with the sources of other mixtures. The authors showed the effectiveness of their approach on synthetic PNL mixtures, synthetic overdetermined mixtures, and two mixed recordings of speech data. As is the current trend in deep learning research, transformer models also show an impressive capability for audio source separation due to their explicit design for sequence modeling \citep{9413901, inproceedingschen2020, simon2022, DBLP:journals/corr/abs-2302-11824}. Mixed audio sources can often be effectively separated with the assumption that the mixture is an approximately linear combination of the sources. However, finding the appropriate unmixing matrix or separation mask for the mixture can still be challenging as the literature shows. The primary focus of this paper is on single-channel non-linear mixtures, which are less explored in the literature, but have the potential to be useful in other domains such as biosignals processing.

Other previously unmentioned domains where deep learning approaches to source separation are in use include finance \citep{e24010055}, chemical analysis \citep{Ando2015ABS, articleduarte2014}, and satellite communications \citep{5406547, 9837851, yang2019blind, 10.3389/frspt.2021.756478, ZHONG20123737}. There are many challenges surrounding source separation and BSS tasks, but data-driven solutions, such as deep learning, are shown to be an effective strategy for solving various problems \citep{10.1016/j.neucom.2023.126895}.

\section{Further Details on Proposed Method} \label{appendix: model details}
\subsection{Implementation and Algorithm}
\begin{figure}[!htb]
  \centering
  \begin{subfigure}[]{0.475\textwidth}
        \centering
        \includegraphics[width=1.0\linewidth]{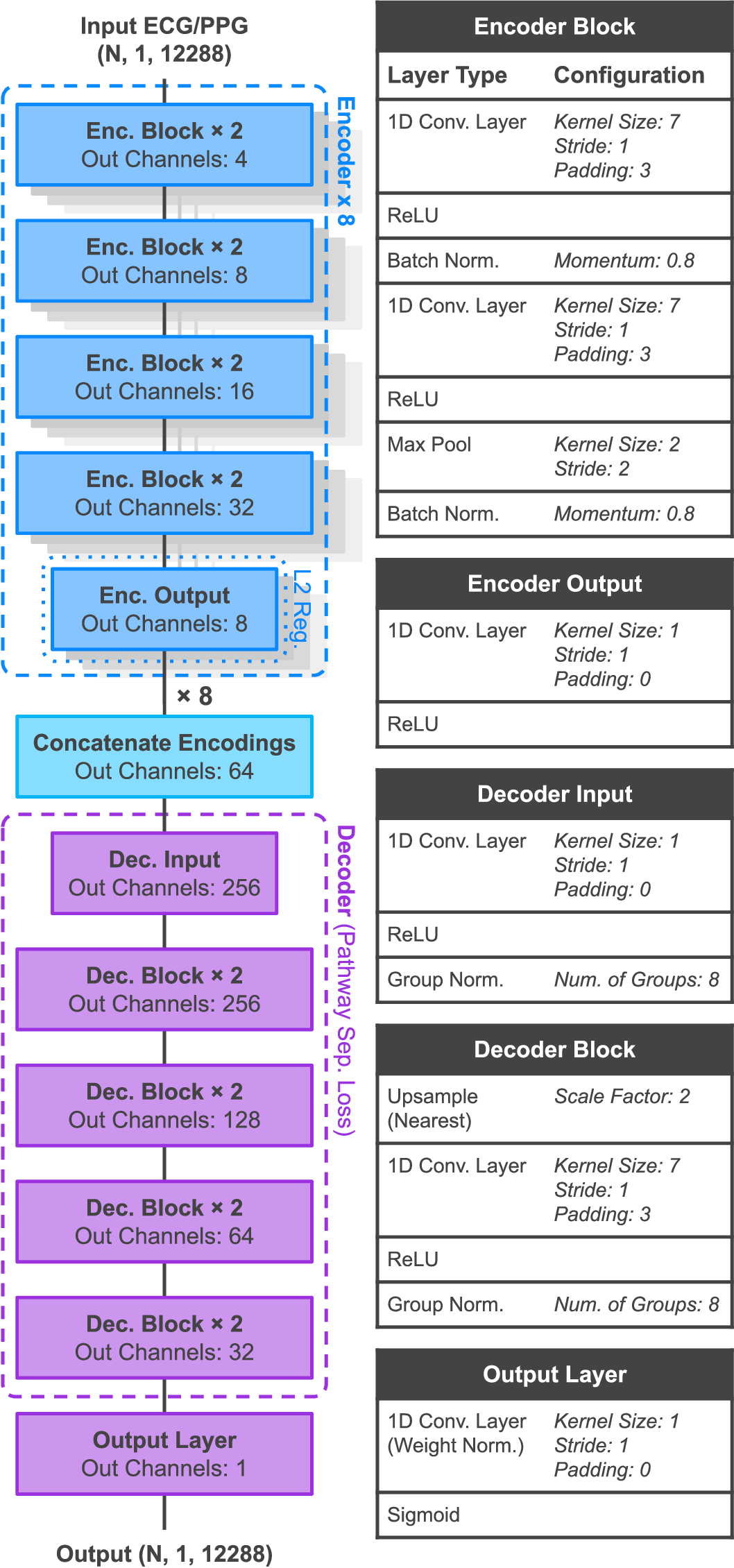}
        \caption{ECG/PPG model experiment specifications.}
        \label{fig: ecg ppg model spec}
  \end{subfigure}\hspace{0.0499\textwidth}%
  \begin{subfigure}[]{0.475\textwidth}
        \centering
        \includegraphics[width=1.0\linewidth]{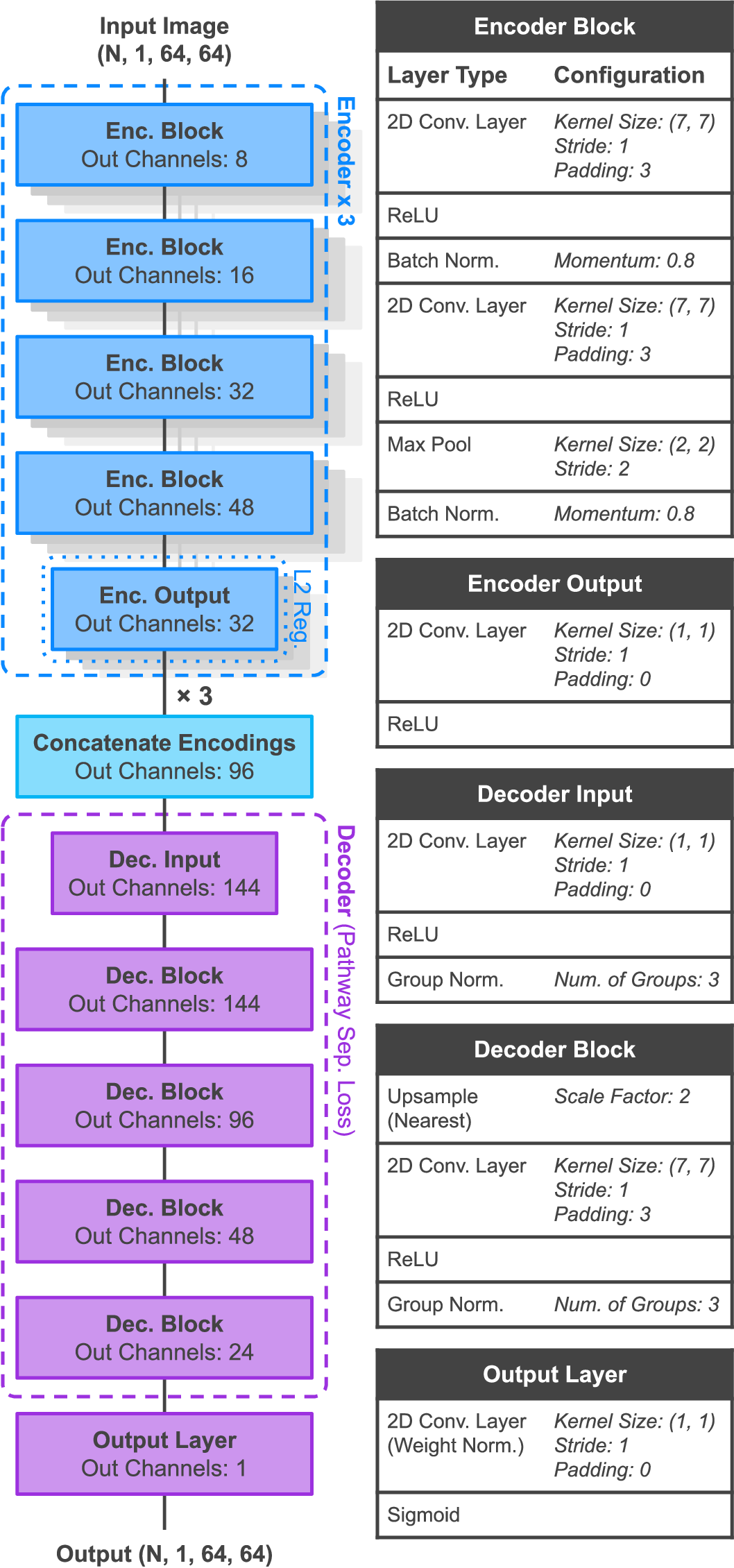}
        \caption{Triangles \& circles dataset experiment model specifications.}
        \label{fig: tri circl model spec}
  \end{subfigure}
  \caption{Model specifications for the ECG/PPG experiments and the triangles \& circles dataset experiments.}
  \label{fig: model specs}
\end{figure}

\begin{algorithm}
\caption{Training loop for proposed method}
    \begin{algorithmic}[1]
        \State $\bm{E}_{\bm{\theta}}$ is comprised of $N$ encoder networks $E_{\theta_n}^n$ parameterized by $\bm{\theta} = \left[\theta_{0}\ \theta_{1} \ \ldots \ \theta_{N}\right]$.
        \State $D_{\phi}$ is the decoder network parameterized by $\phi$.
        \State $x$ is a batch of samples in the dataset $X$.
        \State Choose scaling term(s) $\alpha$ (see \autoref{eq: alpha 2}).
        \For{$x$ in $X$}
            \State $\bm{z} \gets \bm{E}_{\theta}(x)$ \Comment{Get encodings, $\bm{z} = \left[z^{0}\ z^{1}\ \ldots \ z^{N}\right]$.}
            \State $Z \gets z^{0}\oplus z^{1}\oplus \ldots \oplus z^{N}$ \Comment{Concatenate encodings along channel dimension.}
            \State $\hat{x} \gets D_{\phi}(Z)$ \Comment{Get reconstruction.}
            \State $\mathcal{L}_{\text{recon.}} \gets \text{BCE}(x, \hat{x})$
            \For{$z^n$ in $\bm{z}$} \Comment{Calculate $\mathcal{L}_{\text{z}}$}
                \State $\mathcal{L}_{\text{z}} \gets \mathcal{L}_{\text{z}} + \frac{1}{N h} \|z^n\|_2^2$ \Comment{See \autoref{eq: z loss}.}
            \EndFor
            \For{$W$ in $\phi$}
                \For{$B_{i,j}$ in $W$} \Comment{For definition of $W$ and $B_{i,j}$ see \autoref{eq: mixing loss}.}
                    \If{$i \neq j$}
                        \State $\mathcal{L}_{\text{mixing}} \gets \mathcal{L}_{\text{mixing}} + \alpha \|B_{i,j}\|_1$ \Comment{$\alpha$ must be chosen from \autoref{eq: alpha 1} or \autoref{eq: alpha 2}.}
                    \EndIf
                \EndFor
            \EndFor
            \State $Z_\text{zero} \gets \left[0\ 0\ 0\ \ldots \ 0\right]$ \Comment{$Z_\text{zero}$ has same shape as $Z$.}
            \State $\mathcal{L}_{\text{zero recon.}} \gets \text{BCE}(0, D_{\phi}(Z_\text{zero}))$
            \State $\mathcal{L}_{\text{total}} = \mathcal{L}_{\text{recon.}} + \lambda_{\text{mixing}} \mathcal{L}_{\text{mixing}} + \lambda_{\text{zero recon.}} \mathcal{L}_{\text{zero recon.}} + \lambda_{\text{z}} \mathcal{L}_{\text{z}}$
            \State Update $\theta$ and $\phi$ by minimizing $\mathcal{L}_{\text{total}}$ via gradient descent.
        \EndFor
    \end{algorithmic}
\label{algo: train}
\end{algorithm}
All experiments were coded using Python (\path{3.10.6}) using the PyTorch (\path{1.13.1}) machine learning framework, and the PyTorch Lighting (\path{2.0.0}) library which is a high-level interface for PyTorch. In \autoref{fig: model specs}, the exact configuration of the multi-encoder autoencoder structure used for each experiment is shown. The supervised autoencoder models used in our experiments are both very similar to the multi-encoder networks used by the proposed method. The main differences are that only a single encoder is used, the encoding space uses a ReLU activation instead of a linear mapping, and batch normalization is used in the decoder instead of group normalization. The number of layers and all training parameters remain the same. For the supervised autoencoder used in the triangles \& circles experiments the channel dimensions of the single encoder network follow the inverse order of those for the decoder network used by the proposed method. For the ECG and PPG experiments, a slightly smaller channel dimension size was sufficient for achieving strong results. The number of layers remains the same, but the channel dimensions for the encoder are as follows: $[16, 16, 32, 32, 64, 64, 128, 128]$ with the inverse order being used for the decoder and a hidden dimension size of $32$.

\paragraph{Training Procedure}
\autoref{algo: train} shows the basic training loop for the proposed method. In lines 1-2, we construct our architecture which has $N$ encoders $\bm{E}_{\bm{\theta}}$ and a single decoder $D_{\phi}$, parameterized by $\theta$ and $\phi$ respectively. In line 4, the scaling term for the sparse mixing loss $\alpha$ is chosen, two examples of which are shown in \autoref{eq: alpha 2}. From line 5 to line 24, for each batch of samples $x$ in the given dataset, the losses are calculated and the weights are updated. In line 6, the encodings $\bm{z}$ are produced by forward propagating our batch of samples through each encoder $E_{\theta_n}^n$. In lines 7-9, the resulting set of encodings is concatenated along the channel dimension to create a combined representation which is forward propagated through the decoder to produce the reconstructions $\hat{x}$ and the binary cross entropy loss is calculated between the batch of reconstructions $\hat{x}$ and the batch of inputs $x$. In lines 10-12, for each encoding $z^n$ in $\bm{z}$, the L2 norm is calculated and the mean is taken. In lines 13-19, the sparse mixing loss is calculated such that the L1 norm,  scaled by $\alpha$, is calculated for each block $B_{i,j}$ in each layer weight $W$ of the decoder (see \autoref{eq: mixing loss}). In lines 20-21, we forward propagate a second time through the decoder with an all-zero matrix $Z_{zero}$ the same size as our concatenated encodings $Z$, calculating the binary cross entropy between the decoded all-zero encoding and an all-zero matrix the same size as our input. In lines 22-23, the losses are summed to give the total loss $\mathcal{L}_{\text{total}}$, and then the loss is minimized to update the parameters via gradient descent. The described training loop is repeated until convergence or acceptable results are achieved.
\begin{algorithm}
\caption{Inference procedure for proposed method}
    \begin{algorithmic}[1]
        \State $\bm{E}_{\bm{\theta}}$ is comprised of $N$ encoder networks $E_{\theta_n}^n$ parameterized by $\bm{\theta} = \left[\theta_{0}\ \theta_{1} \ \ldots \ \theta_{N}\right]$.
        \State $D_{\phi}$ is the decoder network parameterized by $\phi$.
        \State $x$ is a sample or a batch of samples in the dataset $X$.
        \State $\bm{z} \gets \bm{E}_{\theta}(x)$ \Comment{Get encodings, $\bm{z} = \left[z^{0}\ z^{1}\ \ldots \ z^{N}\right]$.}
        \State Choose encoding $z^n$ to decode.
        \State $Z^n = \bm{0} \oplus \ldots \oplus z^{n}  \oplus \ldots \oplus \bm{0}$ \Comment{$\bm{0}$ is an all-zero matrix with the same shape as $z^n$}
        \State $\hat{s}^n = D_{\phi}(Z^n)$ \Comment{Get source estimation.}
    \end{algorithmic}
\label{algo: inference}
\end{algorithm}
\paragraph{Inference Procedure}
\autoref{algo: inference} describes the general inference procedure for producing source estimations with the proposed method. In line 4, the set of encodings $\bm{z}$ for some input or batch of inputs are produced. In lines 5-6, a single encoding $z^{n}$ (the active encoding) is left for decoding and all other encodings are replaced with all-zero matrices $\bm{0}$ of the same size as their original size. These encodings, keeping the order of the chosen encoding in place, are concatenated along the channel dimension. In line 7, the resulting combined encoding is forward propagated through the decoder $D_{\phi}$ to produce the source estimation $\hat{s}^n$ associated with the encoding $z^n$.
\begin{figure}
  \centering
  \includegraphics[width=0.5\textwidth]{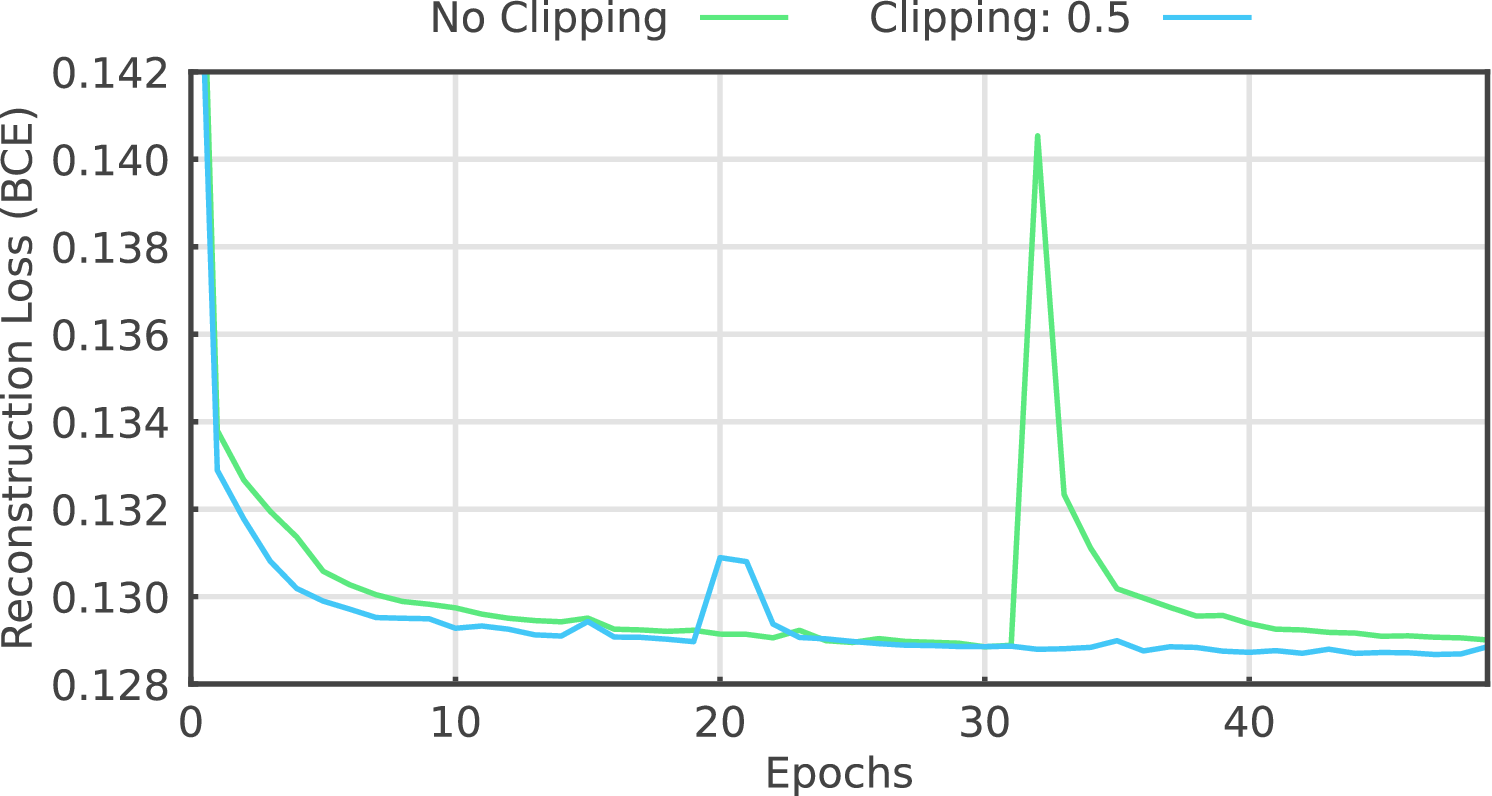}
  \caption{Typical reconstruction loss plot for the first 50 epochs of training on the triangles \& circles dataset with and without gradient clipping.}
  \label{fig: tri_circ_stability}
\end{figure}
\subsection{Training Stability} \label{appendix: training stablility}
During our development of the proposed method, we found that using gradient clipping (for the triangle \& circles experiment we used a value of $0.5$) or, depending on the hyperparameter configuration, applying weight normalization \citep{salimans2016weight} to the output layer (the strategy used for the ECG and PPG experiments) helped to improve training stability, though they are not strictly necessary. Occasionally during training of the proposed method, the model's weights may partially "reset" spontaneously likely as a result of the heavy regularization being applied to the model. The model usually recovers quickly but can take several epochs to fully recover. Gradient clipping significantly helped to mitigate this during our testing of the proposed method on the triangles \& circles dataset. A quick comparison of how the spontaneous resets present themselves in the training reconstruction loss is shown in \autoref{fig: tri_circ_stability}.
\subsection{Indentifiablility and Separability with Non-Linear Mixtures} \label{appendix: Indentifiablility and Separability with Non-Linear Mixtures}
\begin{figure}[H]
  \centering
  \begin{subfigure}[]{0.325\textwidth}
        \centering
            \includegraphics[width=0.8\linewidth]{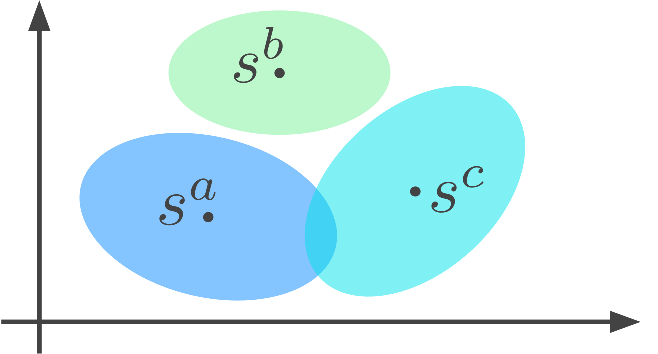}
        \caption{}
        \label{fig: unmixed sub}
  \end{subfigure}\hspace{0.01\textwidth}%
  \begin{subfigure}[]{0.325\textwidth}
        \centering
        \includegraphics[width=0.8\linewidth]{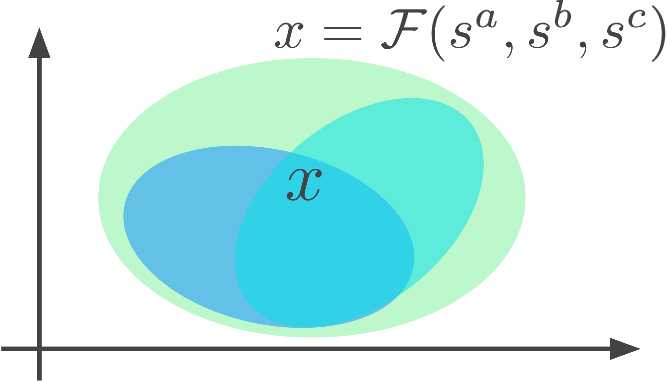}
        \caption{}
        \label{fig: mixed sub}
  \end{subfigure}
  \caption{Illustration showing the mixing of source feature subspaces by a mixing system $\mathcal{F}$.}
  \label{fig: Feature Subspace}
\end{figure}
\autoref{fig: unmixed sub} shows a graph with three well-separated feature subspaces, $s^a$, $s^b$, and $s^c$, that correspond to the sources in a mixture. \autoref{fig: mixed sub} shows a graph where the mixing system $\mathcal{F}$ maps the source feature subspaces into a new feature space where there may be increased ambiguity between the source feature subspaces or in the worst case complete ambiguity. The degree of ambiguity between sources within a mixture also determines how well sources can be recovered, if at all, which is referred to in the literature as separability or identifiability. The issue of separability in non-linear BSS has proven to be a challenging problem due to the additional ambiguities and indeterminacies often introduced by non-linear systems \citep{articleehsandoust, DEVILLE201921}. The issue of separability in non-linear mixtures is unavoidable, and the method proposed in this paper aims to estimate sources even when indeterminacies are present as opposed to finding exact source reconstructions. Source estimations in this scenario may still provide useful representations as shown by our experimental results in \autoref{sub: ppg ecg}. In \autoref{sub: tri circ}, we further showed that the issues of identifiability and separability may be overcome through learning over a distribution of mixtures.

\subsection{Applicability of Alternative Regularization Methods} \label{appendix: regularization}
While not necessary for the types of signals studied in this paper, the loss functions applied to sources proposed by \cite{10285987} (for sparsity, signal intensity, and dependence) could easily be applied to the proposed method by predicting the sources during training and computing each loss. There is a computational disadvantage with the proposed method when applying such losses to predicted sources because each source would require an additional pass through the decoder. Similarly, an adversarial loss applied to source predictions with mixed source encodings could also be used in a similar manner to the method proposed by \cite{Brakel2017LearningIF}. However, the advantage of the proposed method is that the sources can be learned purely through the feature subspace specialization ability of the multiple encoders. Additional priors could be enforced through additional losses on the sources, source encodings or by simply adjusting the weighting applied to the sparse mixing loss (which can be thought of as a type of sparse mixing constraint) and the zero reconstruction loss proposed in this work. The number of proposed losses in the literature for BSS is vast and necessarily application-specific. The application of additional losses to the proposed method may be a promising direction for future research.

\subsection{BSS via Learning from a Distribution} \label{appendix: learning from distribution}
Classical BSS methods like ICA, PCA, and NMF traditionally focus on BSS with a single sample, where the separation happens on a single mixed signal without any kind of training on large amounts of data. In this regard, the proposed approach has a clear disadvantage in that it cannot perform single-sample BSS due to its reliance on learning from a distribution, but this may also be an advantage in cases where single-sample BSS is not viable.

\section{Additional Notes on Triangles \& Circles Experiment}\label{appendix: tri_circ}
\subsection{Optimizer and Hyperparameter Settings} \label{appendix: Optimizer and Hyperparameter Settings}
The Adam optimizer is used for optimization with a learning rate of $1 \times 10^{-3}$, and a step learning rate scheduler is applied, multiplying the current learning rate by $0.1$ every 50 epochs. The model trains for 100 epochs in total, and a batch size of $256$ is used. In addition, a weight decay of $1 \times 10^{-5}$ over all parameters is used. Lastly, three scalars (in addition to the global learning rate) are used to control the contribution of the encoding regularization loss, the sparse mixing loss, and the zero reconstruction loss which are denoted as $\lambda_{\text{mixing}}$, $\lambda_{\text{zero recon.}}$, and $\lambda_{\text{z}}$. For this experiment, $\lambda_{\text{zero recon.}}$ and $\lambda_{\text{z}}$ are set to be $1 \times 10^{-2}$ and $\lambda_{\text{mixing}}$ is set to $5 \times 10^{-1}$.
\begin{gather} 
\mathcal{L}_{\text{total}} = \mathcal{L}_{\text{recon.}} + \lambda_{\text{mixing}} \mathcal{L}_{\text{mixing}} + \lambda_{\text{zero recon.}} \mathcal{L}_{\text{zero recon.}} + \lambda_{\text{z}} \mathcal{L}_{\text{z}}
\end{gather}
\subsection{Selection of the Distortion Kernel} \label{appendix: selection of distortion kernel}
The purpose of convolving the sigmoid PNL mixture by the proposed distortion kernel is to partially discard spatial and shape information in the sigmoid PNL mixture. The convolution operation makes estimating an inverse mapping to the mixing system more difficult due to the introduced indeterminacies. The random horizontal flipping of the kernel means that our mixing system also varies from sample to sample further adding to the difficulty. The distortion kernel is designed to be asymmetric along the vertical line so that horizontal flipping creates a new kernel that mirrors the original.

Further, because of the sigmoid activation used at the output of the network, using a distortion kernel after the sigmoid-based PNL mixing prevents the trivial solution in which the decoder reconstructs the sources prior to the output layer, and then the output layer simply applies linear mixing followed by the final sigmoid function (this is confirmed by the additional experiments in \autoref{appendix: generalizing to the mixing system}). Additionally, the kernel includes patches of all zeros that destroy information about the edges of the shape. While the exact kernel used is not of great importance for the discussions in this paper, the kernel used in our experiments is provided below for reproducibility.
\begin{gather}
    \text{Distort. Kernel}=\begin{bmatrix}  
    1.0 & 1.0 & 0.0 & 0.0 & 0.0 \\
    0.0 & 0.0 & 0.5 & 1.0 & 0.5 \\
    0.0 & 0.0 & 0.0 & 0.5 & 1.0 \\
    0.0 & 0.0 & 0.5 & 1.0 & 0.5 \\
    1.0 & 1.0 & 0.0 & 0.0 & 0.0 
    \end{bmatrix}
\end{gather} 
\subsection{Generalizating to the Mixing System} \label{appendix: generalizing to the mixing system}
\begin{figure}[H]
  \centering
  \begin{subfigure}[]{0.35\textwidth}
        \centering
            \includegraphics[width=1.0\linewidth]{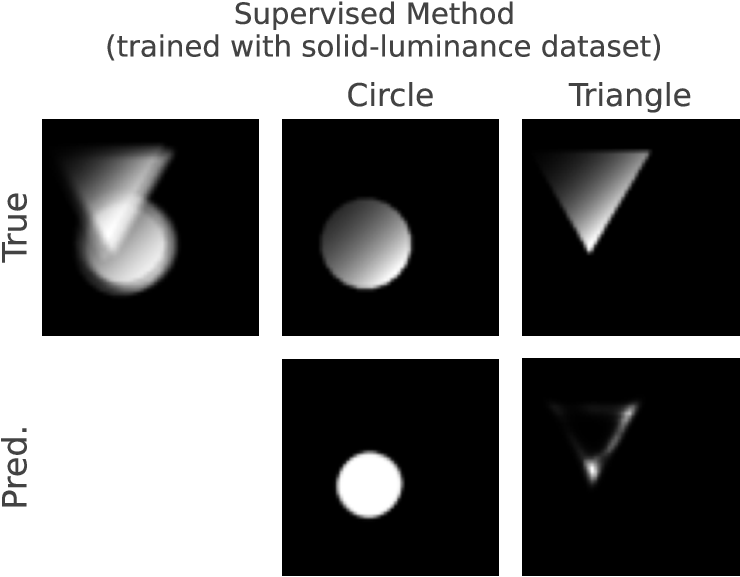}
        \caption{}
        \label{fig: tri circ gen supervised}
  \end{subfigure}\hspace{0.1\textwidth}%
  \begin{subfigure}[]{0.35\textwidth}
        \centering
        \includegraphics[width=1.0\linewidth]{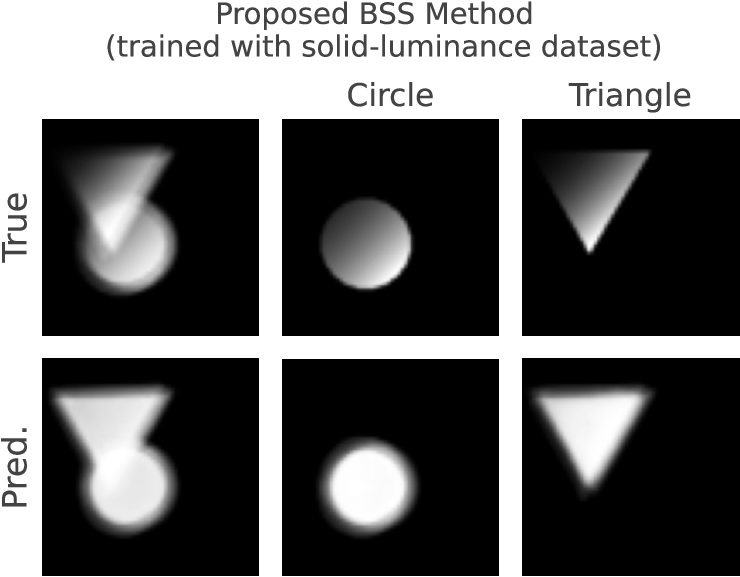}
        \caption{}
        \label{fig: tri circ gen bss}
  \end{subfigure}
  \caption{Comparison of a supervised approach and the proposed BSS approach for reproducing test-set sources when they are perturbed after the training. In this case, a gradient is applied to the sources.}
  \label{fig: tri circ supervised vs unsupervised}
\end{figure}
In \autoref{fig: tri circ supervised vs unsupervised} the ability of the proposed method to generalize to the mixing system is exemplified. Before applying the mixing system to the sources a gradient mask is applied to the triangle and circle sources. Then the proposed approach and the supervised approach trained on the original dataset are used for source predictions. While the supervised approach often fails to accurately predict the sources, the proposed BSS method performs reasonably well (though the network is unable to preserve the gradient applied to the sources). This is evidence that the proposed method is trained to learn the mixing system itself and the structure of the shapes, while the supervised approach needs only to learn how to reproduce sources given the holistic structural context of the image.
\begin{figure}[H]
    \centering
  \begin{subfigure}[]{0.35\textwidth}
        \centering
            \includegraphics[width=0.9\linewidth]{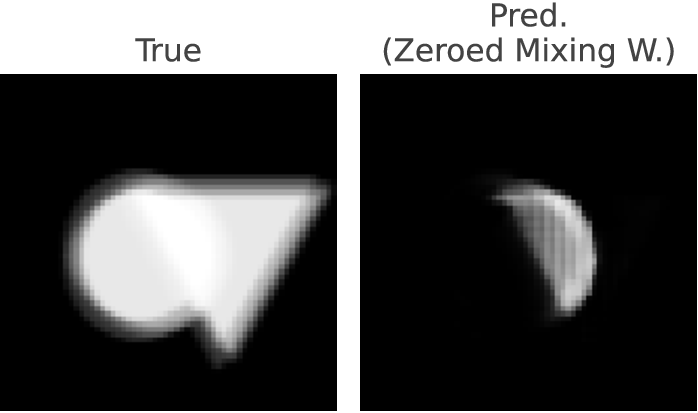}
            \caption{}
  \end{subfigure}\hspace{0.1\textwidth}%
  \begin{subfigure}[]{0.35\textwidth}
        \centering
        \includegraphics[width=0.9\linewidth]{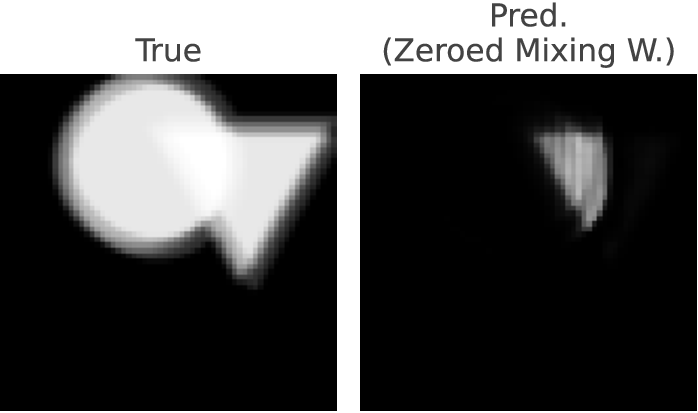}
        \caption{}
  \end{subfigure}
  \caption{The predicted mixture by the proposed method after zeroing the off-diagonal blocks in the weights of the decoder.}
  \label{fig: zeroed}
\end{figure}

Further, in \autoref{fig: zeroed} we show that by first zeroing the off-diagonal blocks for each weight in the decoder, the network is \emph{only} able to predict overlapped regions of the triangle and circle shapes when predicting the mixture. The off-diagonal blocks within the decoder weights model the non-linear relationship between shapes. Referencing \autoref{eq: tri circ mxing} and ignoring the distortion kernel, while it is true that for elements in overlapping regions (with values 1.0), the transformation is effectively linear, this is not true for non-overlapping regions (with values 0.5). 
\begin{gather}
\frac{1}{2}(1.0 + 1.0) = \text{Scaled}(\sigma(\frac{6}{2}(1.0 + 1.0)))\\
\frac{1}{2}(1.0 + 0.0) \neq \text{Scaled}(\sigma(\frac{6}{2}(1.0 + 0.0)))
\end{gather}

where $\text{Scaled}$ is a min-max function applied with respect to all of the elements in the PNL sigmoid mixture which would have a minimum value of $0.0$ and a maximum value of approximately $0.9975$ before scaling. Thus, while zeroing the off-diagonal blocks within the weights of the decoder, the model fails to predict the non-overlapping and non-linearly mixed regions of the mixture.

\subsection{Application of Common BSS Approaches} \label{appendix: Application of Common BSS Approaches}
Common BSS algorithms were originally designed specifically for time-series data. In some cases, they can be adapted for image data, which contains an additional spatial dimension \citep{engproc2022014020, 5470083, XIANCHUAN2013288}, particularly in the case of linear mixtures. The triangle \& circles dataset has unique statistical properties that make it a valuable tool for demonstrating the proposed method's ability to learn separation via the distribution of the dataset, but less ideal for the common assumptions made by many BSS approaches. For starters, the triangle and circle sources have no distinguishing characteristics in their amplitudes, position, or scale. The distribution of both sources in terms of scale and position is uniform, and the amplitudes are approximately binary. The only distinguishing features between the sources are their shapes (e.g. straight edges for the triangle sources and round edges for the circle sources). Additionally, because there is only one observation in the mixture (i.e. a single-channel mixture), point-wise information is insufficient for separating these sources on a per-sample basis. Thus separation requires a structural understanding of the objects themselves which can only be learned by either knowing information about the sources in advance or by learning from and generalizing to the entirety of the dataset.

\section{Additional Notes on ECG and PPG Experiments}\label{appendix: ecg ppg}
\begin{figure}
  \centering
  \includegraphics[width=0.8\textwidth]{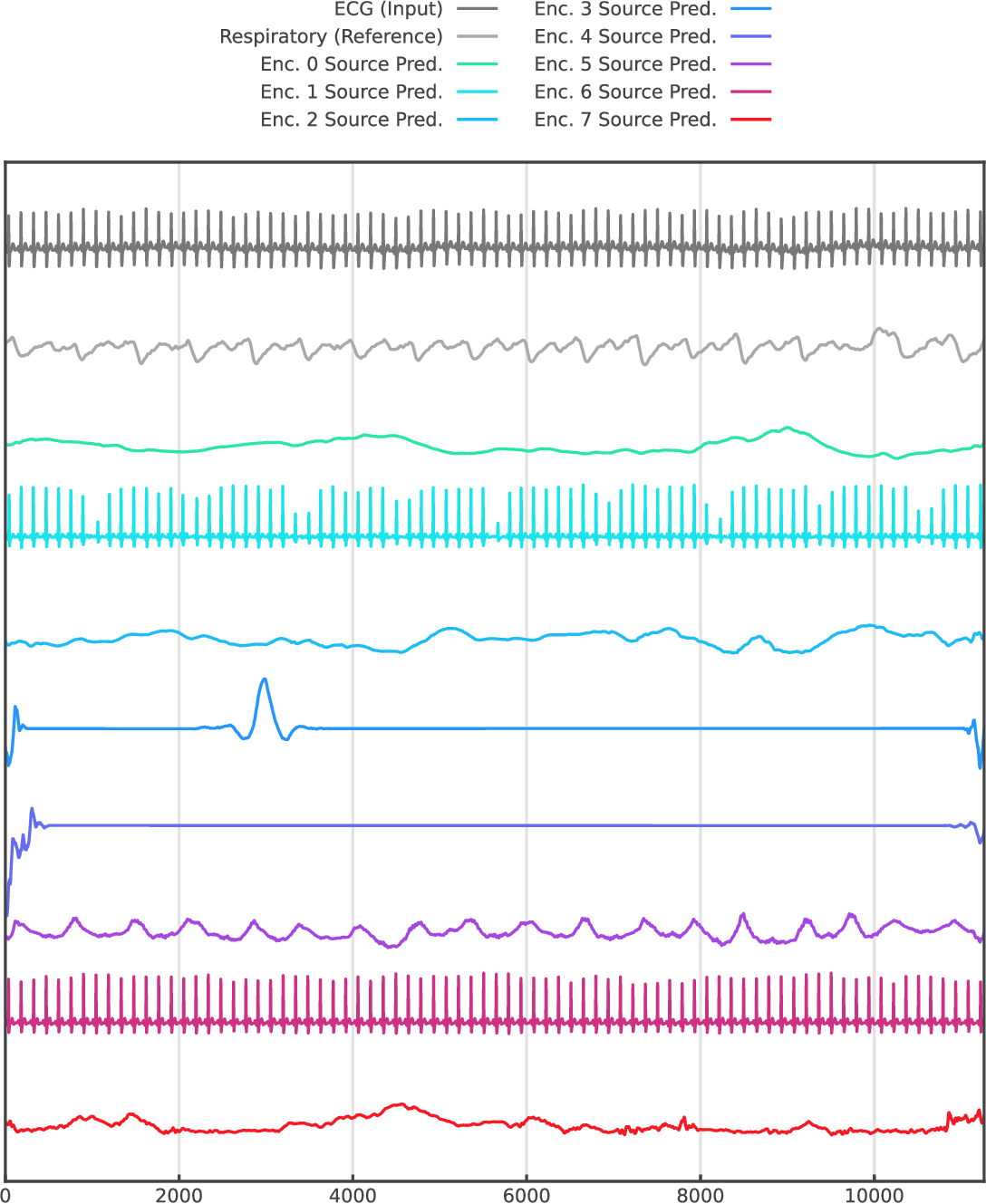}
  \caption{Figure showing all of the extracted sources using the proposed method for the ECG experiment.}
  \label{fig: ecg sources}
\end{figure}

\autoref{fig: ecg sources} shows all eight sources predicted for an input ECG signal using the proposed method. Baseline wander (\emph{Enc. 0} and \emph{Enc. 3}), noise (\emph{Enc. 7}), cardiovascular/heart (\emph{Enc. 6}), and respiratory sources (\emph{Enc. 5}) can all be seen. An example of a dead encoder source prediction can also be seen among the source predictions (\emph{Enc. 4}). It is not unusual for the baseline wander sources to correlate with the respiratory source, but because ECG and PPG signals are affected by three types of respiratory-induced modulation, they predict much less accurate representations of the respiratory signal overall. As for the other source predictions, it is much more difficult to decipher their meaning without a deeper analysis which is beyond the scope of this work.
\begin{figure}[H]
  \centering
  \begin{subfigure}[]{1.0\linewidth}
        \centering
        \includegraphics[width=1.0\textwidth]{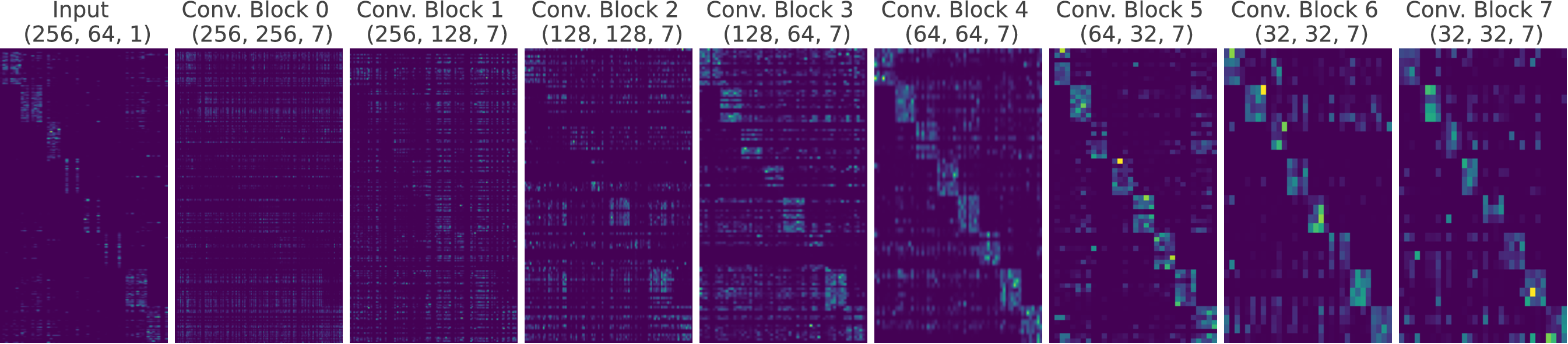}
        \caption{Absolute values of ECG model decoder weights.}
        \label{fig: ecg weights}
  \end{subfigure}
  \begin{subfigure}[]{1.0\linewidth}
        \centering
        \includegraphics[width=1.0\textwidth]{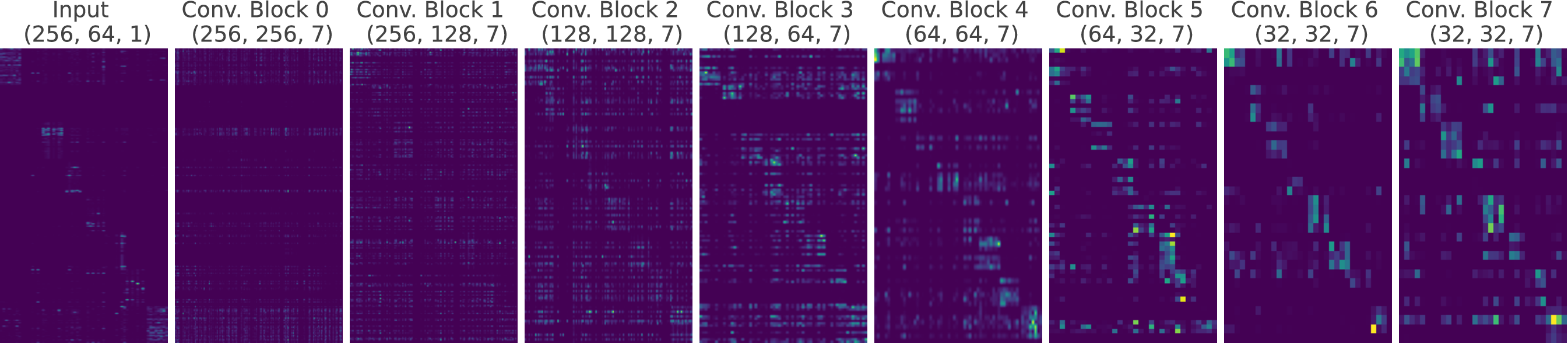}
        \caption{Absolute values of PPG model decoder weights.}
        \label{fig: ppg weights}
  \end{subfigure}
  \caption{The absolute values of the decoder weights summed along the spatial dimension for the PPG and ECG models showing the application of the \emph{sparse mixing loss}.}
  \label{fig: ecg_ppg_weights}
\end{figure}
The absolute values of the decoder weights summed along the spatial dimension for the PPG and ECG models are shown in \autoref{fig: ecg_ppg_weights}. Due to the complex non-linear relationships that occur between sources in a PPG or ECG signal, a lower penalty is applied to the sparse mixing loss, and the alternative $\alpha$ scalar \eqref{eq: alpha 2} is applied. Thus, the weights within the off-diagonal blocks have generally higher values in comparison to the model trained for the triangles \& circles dataset experiment.
\subsection{Optimizer and Hyperparameter Settings} \label{appendix: Optimizer and Hyperparameter Settings ecg_ppg}
The Adam optimizer with a learning rate of $1 \times 10^{-3}$ is used for optimizing the model weights. Additionally, a weight decay of $1 \times 10^{-6}$ is applied over all parameters. The scalars for the additional regularization terms, $\lambda_{\text{mixing}}$, $\lambda_{\text{zero recon.}}$, and $\lambda_{\text{z}}$, are set to $1 \times 10^{-3}$, $1 \times 10^{-2}$, and $1 \times 10^{-2}$, respectively. The batch size is set to $256$ segments. The models for both PPG and ECG are trained for $50$ epochs, and then the best-performing model from these runs is chosen for the final evaluation on the test set. 
\subsection{Efficacy of Common BSS Approaches} \label{appendix: efficacy of common bss approaches}
For comparison, we tested several common BSS approaches on the task of separating respiratory signals from unprocessed ECG and PPG signals: NMF \citep{Lee1999}, SparsePCA \citep{pmlr-v9-jenatton10a, pmlr-v28-kuleshov13}, KernelPCA \citep{10.1007/BFb0020217}, FastICA \citep{NIPS1996_dfd7468a}, and adversarial networks for Non-Linear ICA \citep{Brakel2017LearningIF}. All methods failed to produce any meaningful respiratory signal due to the core assumptions that these methods make such as linear and stationary mixing which are not ideal assumptions for ECG and PPG signals when extracting a respiratory source as discussed in \autoref{sub: ppg ecg}. However, these more common BSS methods may be suitable for removing noise as it can often be considered linearly separable.

\subsection{Performance of Heuristic Approaches}
The heuristic approaches included in \autoref{ecg-ppg-table} were shown in a previous work to have reasonable performance in estimating respiratory signals from ECG and PPG signals \citep{iet:/content/conferences/10.1049/cp.2015.1654}. Unlike the final dataset used in this paper, the authors put a lot of care into additional filtering for the ECG, PPG, and nasal flow signals and into removing low-quality signals from the dataset. During PSG studies, it is not uncommon for patients, especially those with sleep apnea, to move in their sleep. For the subset of the MESA dataset used in this paper, no additional filtering is applied to the ECG or PPG signals, and signals are only removed if the average heart rate is simply detected to be outside the range of $40$-$180$ beats per minute. The inclusion of low-quality data explains the discrepancy in performance compared to prior works and highlights the robustness of the proposed method to low-quality ECG and PPG signals. 

Some of the heuristic approaches combine BSS approaches with filtering or heuristic algorithms \citep{6144719, s20113238}, but many common BSS approaches alone are insufficient for separating respiratory signals from ECG and PPG. From one perspective, the proposed method could be viewed as a self-supervised end-to-end version of a filtering-based method, with learned separation, because of the view of convolutional layers as learned filters.

\end{document}